\documentclass[12pt,preprint]{aastex}
\def\ltsima{$\;\buildrel < \over \sim \;$}
\def\simlt{\lower.5ex \hbox{\ltsima}}
\def\gtsima{$\;\buildrel > \over \sim \;$}
\def\simgt{\lower.5ex \hbox{\gtsima}}
\usepackage[dvips]{color}

\shorttitle{Interstellar halogen chemistry}
\shortauthors{Neufeld \& Wolfire}

\begin{document}

\title{The chemistry of interstellar molecules \\
 containing the halogen elements}
\author{David A. Neufeld\altaffilmark{1} and Mark G. Wolfire\altaffilmark{2}}
\altaffiltext{1}{Department of Physics and Astronomy, Johns Hopkins University,
3400 North Charles Street, Baltimore, MD 21218; neufeld@pha.jhu.edu}
\altaffiltext{2}{Department of Astronomy, University of Maryland, College Park, MD 20742; mwolfire@astro.umd.edu}

\begin{abstract}

Although they are only minor constituents of the interstellar medium, halogen-containing molecules are of special interest because of their unique thermochemistry.  Here, we present a theoretical study of the chemistry of interstellar molecules containing the halogen elements chlorine and fluorine.  We have modeled both diffuse and dense molecular clouds, making use of updated estimates for the rates of several key chemical processes.  We present predictions for the abundances of the three halogen molecules that have been detected to date in the interstellar medium:  HF, CF$^+$ and HCl.  As in our previous study of fluorine-bearing interstellar molecules, we predict HF to be the dominant gas-phase reservoir of fluorine within both diffuse and dense molecular clouds; we expect the {\it Herschel Space Observatory} to detect widespread absorption in the HF $J=1-0$ transition.  Our updated model now overpredicts the CF$^+$ abundance by a factor $\simgt 10$ relative to observations of the Orion Bar; this discrepancy has widened because we now adopt a  
laboratory measurement of the CF$^+$ dissociative recombination rate that is smaller than the estimate we adopted previously.  This disagreement suggests that the reaction of C$^+$ with HF proceeds more slowly than the capture rate assumed in our model; a laboratory measurement of this reaction rate would be very desirable.  Our model predicts diffuse cloud HCl abundances that are similar to those predicted previously and detected tentatively toward $\zeta$~Oph. Two additional species are potentially detectable from photodissociation regions: the  $\rm H_2Cl^+$ and $\rm HCl^+$ molecular ions.  Ortho-$\rm H_2Cl^+$ has its lowest-lying transition in the millimeter spectral region observable from the ground, and the lowest rotational transition of HCl$^+$ is observable with {\it Herschel}'s HIFI instrument.

\end{abstract}

\keywords{ISM: Molecules --- ISM: Abundances --- ISM: Clouds -- molecular processes -- infrared: ISM -- submillimeter}

\section{Introduction: the unique thermochemistry of fluorine and chlorine}

Of the $\sim$ 150 distinct molecules detected thus far in interstellar clouds and circumstellar outflows, over  90$\%$ contain only hydrogen and/or elements in groups IVA, VA, and VIA\footnote{Here we use the traditional designations used in the United States.  The current IUPAC designations for groups IVA, VA, and VIA are respectively 14, 15, and 16; those for groups IA, IIA, IIIA and VIIA are 1, 2, 13, and 17}
of the periodic table: viz.\ carbon, nitrogen, oxygen, silicon, phosphorus, and sulfur.  Of the remaining 12 detected molecules, eight contain elements in groups IA, IIA and IIIA (viz.\ sodium, potassium, magnesium and aluminum) and are found exclusively in circumstellar envelopes (e.g.\ Ziurys 1996, and references therein) with abundances that are set in cool stellar photospheres; one contains iron (FeO, detected toward the Sgr B2 interstellar gas cloud by Walmsley et al.\ 2002); and the remaining 3 are interstellar molecules containing elements of the halogen group (VIIA): HCl, HF, and CF$^+$. 

Although they are only minor constituents of the interstellar medium, halogen-containing molecules are of special interest because of their unique thermochemistry.  The basic behavior of any atom in the interstellar medium is determined by two or three key thermochemical considerations.  First, the first ionization potential (IP) determines whether the atom will be neutral or ionized within diffuse molecular clouds: an atom X with IP(X) $>$ IP(H) = 13.6 eV will be shielded by atomic hydrogen from the interstellar radiation field (ISRF) and will be predominantly neutral, while an atom with IP(X) $<$ IP(H) will be predominantly ionized.  Second, the dissociation energy of the neutral hydride, $D_0$(HX), determines whether X can react exothermically with H$_2$: atoms for which $D_0$(HX) $>$ $D_0$(H$_2$) = 4.48 eV can react exothermically with H$_2$ to form HX, while atoms with  $D_0$(HX) $ < D_0$(H$_2$) cannot.  Third (and of importance mainly for atoms with IP $<$ 13.6 eV),  the dissociation energy of the hydride molecular ion, $D_0$(HX$^+$), similarly determines whether the ion X$^+$ can react exothermically with H$_2$ to form HX$^+$.  Fluorine and chlorine\footnote{In this paper we combine our attention to Cl and F; bromine and iodine have much smaller cosmic abundances than chlorine and fluorine, and molecules containing bromine and iodine are unlikely to be detectable with current technology.}
are unique in regard to the second and third criteria, each in a different way: F is the only atom -- and Cl$^+$ is the only ion with an appearance potential $<$ 13.6 eV -- that can react exothermically with H$_2$ to form a diatomic hydride.  

The uniqueness of these elements are presented graphically in Figure 1, which shows $D_0$(HX) for each atom X in the first three rows of the periodic table.  The molecular binding energy of HX is largest for fluorine and decreases, nearly monotonically, as one moves leftwards and downwards in the periodic table.  In Figure 2, $D_0$(HX) and $D_0$(HX$^+$) are shown for all the elements of which interstellar molecules have been detected.  Red squares apply to atoms with  IP $<$ IP(H) and blue squares to those with IP $>$ IP(H).  With the exception of the halogen elements, all the elements fall into two classes: 

(1) {\it Oxygen and nitrogen, with IP $>$ IP(H), $D_0$(HX) $<$ $D_0$(H$_2$) and $D_0$(HX$^+$) $>$ $D_0$(H$_2$).} These elements are predominantly neutral in the diffuse ISM.  Their atoms do not react with H$_2$ except at elevated temperatures that can drive endothermic reactions, and their chemistry is driven by cosmic ray ionization.

(2) {\it Carbon, silicon, sulfur, and phosphorus, with IP $<$ IP(H), $D_0$(HX) $<$ $D_0$(H$_2$) and $D_0$(HX$^+$) $<$ $D_0$(H$_2$).}  These elements are predominantly ionized in the diffuse ISM.  Their ions do not react with H$_2$ except at elevated temperatures that can drive endothermic reactions.

Chlorine and fluorine form two additional classes with one member each:

(3) {\it Fluorine, with IP $>$ IP(H) and $D_0$(HX) $>$ $D_0$(H$_2$).}  Fluorine is predominantly neutral in the diffuse ISM, and can react exothermically with H$_2$ to form HF.

(4) {\it Chlorine, with IP $<$ IP(H), $D_0$(HX) $<$ $D_0$(H$_2$) and $D_0$(HX$^+$) $>$ $D_0$(H$_2$).}  Chlorine is predominantly ionized in atomic clouds.  Cl$^+$ can react exothermically with H$_2$ to form HCl$^+$.

The chemistry of interstellar fluorine has been considered in a previous study by Neufeld, Wolfire \& Schilke (2005; hereafter NWS05), while that of chlorine has been discussed in several studies, most recently those of Schilke, Wang \& Phillips (1995) and Amin (1996).  In the present paper, we extend and update these previous investigations to account for recent developments in experimental and theoretical studies of relevant chemical processes.  Our study is also motivated by the prospects for upcoming observations of halogen-containing molecules, particularly with the {\it Herschel Space Observatory}.

\section{Model}

\subsection{Chemical network for chlorine-bearing species}

The key chemical pathways leading to the formation and destruction of chlorine-bearing molecules have been elucidated in the papers of Jura (1974), Dalgarno et al.\ (1974),
van Dishoeck \& Black (1986; hereafter vDB86), Blake, Anicich \& Huntress (1986; hereafter BAH86), Schilke, Phillips \& Wang (1995), Federman et al.\ (1995), and Amin (1996). 
In diffuse molecular clouds, the chemical network is initiated by reaction of Cl$^+$ with H$_2$:

\begin{equation}
\rm Cl^+ + H_2 \rightarrow HCl^+ + H,
\end{equation}
which is followed by dissociative recombination of HCl$^+$ or a further H atom abstraction reaction to form H$_2$Cl$^+$:
\begin{equation}
\rm HCl^+ + H_2 \rightarrow H_2Cl^+ + H
\end{equation}
The $\rm H_2Cl^+$ molecular ion undergoes dissociative recombination to form either HCl or Cl
\begin{equation}
\rm H_2Cl^+ + e  \rightarrow HCl + H
\end{equation}
\begin{equation}
\rm H_2Cl^+ + e \rightarrow Cl + products 
\end{equation}
The branching ratio for dissociative recombination of H$_2$Cl$^+$ has not yet been measured, but has been assumed in previous studies to be only 10$\%$ for production of HCl (reaction 3).  This assumption was largely motivated by the small HCl abundances observed in diffuse clouds (see \S 4 below.) 
HCl is destroyed by photodissociation, by photoionization and by reactions with He$^+$, H$_3^+$ and C$^+$.

In regions where H$_2$ is present, reactions (1), (2) and (4) can rapidly convert Cl$^+$ to Cl.  The rate at which the reaction network is initiated (reaction 1) is therefore limited by the availability of Cl$^+$ ions.  These are produced primarily by the photoionization of Cl, following the absorption of ultraviolet photons in the $912 - 958 \AA$
wavelength range.  Near cloud surfaces, the UV radiation field is dominated by starlight (i.e.\ by the interstellar radiation field (ISRF) or nearby OB associations), whereas in the deep interiors of dense clouds, the radiation largely comprises H$_2$ Lyman and Werner band emissions excited by the secondary electrons produced by cosmic rays.
Here, proton transfer from H$_3^+$ to Cl, producing HCl$^+$, can also initiate the formation of Cl-bearing molecules.

Taking as our starting point the reaction list and compilation of rate coefficients in the UDFA (formerly UMIST) rate file (Woodall et al.\ 2007), we have reviewed the rate coefficients critically and implemented the following changes:

1) For reaction of HCl with He$^+$, H$_3^+$ and C$^+$, we have used the statistical adiabatic capture model (SACM; Troe 1985, 1987, 1996; see NWS05 for details of exactly how this was implemented) to estimate the temperature dependence of the capture rate.  Over the temperature range 10 -- 300 K, we found the capture rate to be proportional to $T^{-0.202}$ for rotationally-cold HCl; accordingly we adopted a $T^{-0.202}$ dependence for the assumed reaction rate.  Similar considerations lead to a $T^{-0.194}$ temperature dependence for the rate coefficient for the reaction of H$_2$O with H$_2$Cl$^+$.  We also adjusted the T = 300 K rate coefficients in the UDFA compilation, which were based upon measurements of the thermal rate coefficients, to account for the fact that HCl is primarily in J = 0 (i.e.\ rotationally-cold) under the astrophysical conditions of typical interest; here again, we used the SACM model to estimate the necessary correction factors.  

2) As in NWS05, we adopted a rate coefficient of $2 \times 10^{-7} (T/300\,{\rm K})^{-1/2}\,\rm cm^3 \,s^{-1}$ 
for the dissociative recombination of diatomic molecular ions (in the absence of experimentally-determined values).
For the dissociative recombination of H$_2$Cl$^+$, we adopted a rate coefficient of $1.2 \times 10^{-7} (T/300\,{\rm K})^{-0.85}\,\rm cm^3 \,s^{-1}$, based upon preliminary results from a recent laboratory measurement (Geppert \& Hamberg 2009).  Because the branching ratios for the production of HCl and Cl were not determined in this experiment, we followed previous authors in adopting - as our standard assumption - a branching ratio of 0.1 for the production of HCl.  As discussed in \S4.2 below, we also investigated how the model predictions depend upon the assumed value of this branching ratio.

3) We adopted the rate coefficients of Pradhan \& Dalgarno (1994) for charge transfer from H$^+$ to Cl, and for the reverse reaction (from Cl$^+$ to H).  These reactions were not included in the UDFA compilation, but had been considered by BAH86, who assumed a large rate coefficient of 10$^{-9} \rm \,cm^3 s^{-1}$ for the reaction of H$^+$ with Cl and concluded that this process was among the subset of most significant reactions.  The calculations of Pradhan \& Dalgarno (1994), however, imply rate coefficients that are smaller than those guessed by BAH86 by factors of $\sim 20$ and $\sim 250$ at temperatures of 300 K and 100 K respectively.  

4) We have also included the reaction of Cl with H$_2$ to form HCl and H, which is slightly endothermic (by $\sim 0.05$~eV), along with its reverse reaction.   This reaction, which is not included in the UDFA compilation, has been the subject of intensive investigation, both experimental (e.g. Kumaran, Lim \& Michael 1994
; hereafter KLM94) and theoretical (e.g.  Manthe, Capecchi \& Werner 2004
).  Here we adopted the rate coefficients recommended by KLM94 (their equations 10, 11 and 6), which were obtained by a fit to their own measurements together with those obtained in previous studies.

5)  In computing revised estimates of the HCl photodissociation and photoionization rates, we have used the photoabsorption cross-sections obtained by Brion, Dyke \& Cooper (2005)
by means of dipole (e, e) spectroscopy using an electron energy loss spectrometer (EELS).   This technique is believed to offer significant advantages over direct UV absorption measurements for the case of strong absorption bands in which line saturation is important (Chan, Cooper \& Brion 1991)
For example, in the case of excitation to the M$^1\Pi$ state by photons near 110.3 nm, Brion et al.\ obtained an oscillator strength that was a factor of 4 larger than that obtained in the earlier UV absorption experiments of Nee, Suto \& Lee (1986).
The continuous absorption cross-sections obtained by Brion et al.\ (2005) are in good agreement with previous UV absorption measurements (e.g. Cheng et al.\ 2002).

To determine the photodissociation rate implied by the measured photoabsorption cross-sections, the photodissociation probability must also be known.  For radiation longward of 115 nm, the fluorescence yield has been measured (Nee et al.\ 1986) and found to be small, so the photodissociation probability is very close to unity.  For absorption in the 105.5 -- 115 nm region, the fluorescence at wavelengths longer than 115 nm was observed to be weak, but the experiment of Nee et al.\ (1986) did not exclude the possibility of strong fluorescent emission in the 105.5 -- 115 nm region.  For the 97.3 -- 105.5 nm spectral region, we are unaware of any experimental constraints on the fluorescent yield.  In computing the photodissociation rate, we have assumed that the dissociation probability is unity\footnote{At the other extreme, if we assume the photodissociation yield to be zero for radiation shortward of 115 nm, the unshielded photodissociation rate is $0.77 \times 10^{-9} \rm \, s^{-1}$.  This therefore constitutes a lower limit on the true value, and is in good agreement with the corresponding minimum value given by vDvHD.} for all radiation longward of the ionization threshold at $\sim 97.3$~nm.  This leads to a photodissociation rate of $1.7 \times 10^{-9} \, \rm s^{-1}$ for an unshielded molecule exposed to the mean interstellar radiation field given by Draine (1978), a value that  is $ \sim 70\% $ higher than that given by van Dishoeck, van Hemert \&  Dalgarno (1982; hereafter vDvHD).  This difference is largely accounted for by the fact that the total oscillator strength shortward of 129 nm is now known to exceed significantly the simple estimate given by vDvHD.  We obtain an HCl photoionization rate of $1.0 \times 10^{-10}\rm s^{-1}$.

In considering the effect of shielding upon the HCl photodissociation rate, we used a method similar to that adopted by NWS05.  Here, we modeled the case of a semi-infinite slab that is illuminated isotropically, using the photodissociation region (PDR) model of Le Bourlot et al.\ 1993 and Le Petit et al.\ 2006 to determine how the UV radiation field varies with position.  We assumed the 
photodissociation rate to diminish as $E_2(kA_V)$, where $A_V$ is the
visual extinction in magnitudes behind the slab surface, $E_2$ is
the exponential integral of order 2, and $k$ is an adjustable parameter that we varied in order to obtain the best fit to the predicted photodissociation rate.  The UV dust albedo assumed here was 0.32, the forward scattering function was 0.73
(Li \& Draine 2001), and the best fit was obtained for $k=2.1$. 

6) Our estimate of the Cl photoionization rate was based upon measurements of the ionization cross-section by Ruscic \& Berkowitz (1983
), which are in good agreement with theoretical calculations of Brown, Carter \& Kelly (1980
).  Two autoionizing resonances are present at wavelengths longer than the Lyman limit, but make a negligible contribution to the overall photoionization rate.  We derived a photoionization rate of $4.8 \times 10^{-11}\rm s^{-1}$ for an unshielded chlorine atom exposed to the mean interstellar radiation field given by Draine (1978), a value that lies a factor of $\sim 2$ below that adopted by van Dishoeck (1988).

Because the IP of atomic chlorine is only slightly less than that of H, there is only a narrow wavelength range, 912 -- 958$\AA$, over which photoionization can occur.  Radiation in this spectral region is subject not only to absorption by dust, but also to absorption by the H$_2$ Lyman and Werner bands and the HI Lyman lines.  While the effect of HI absorption is small, that of H$_2$ absorption can be significant. We have considered dust and H$_2$ absorption independently, by running the Le Petit
et al.\ (2006) PDR model both with and without H$_2$ absorption.  We have found that the reduction in the Cl photoionization rate can be represented to good approximation by a factor \begin{equation}
E_2(3.6\,A_V) \times {\exp(-N({\rm H}_2)/4.28 \times 10^{21} {\rm cm}^{-2}) \over 
1 + \sqrt {N({\rm H}_2)/4.9 \times 10^{20} {\rm cm}^{-2}} }
\end{equation}
where $N({\rm H}_2)$ is the column density of H$_2$ to the cloud surface.

The photoionization of HCl occurs over a similarly narrow range of wavelengths.  For that process, an identical treatment yields a reduction factor
\begin{equation}
E_2(3.45\,A_V) \times {\exp(-N({\rm H}_2)/3.5 \times 10^{21} {\rm cm}^{-2}) \over (T/300\,{\rm K})
1 + \sqrt {N({\rm H}_2)/3.2 \times 10^{20} {\rm cm}^{-2}} }
\end{equation}

7) Using the cross-sections cited in (5) and (6) above, we have computed the rates of HCl photodissociation, HCl photoionization, and Cl photoionization resulting from H$_2$ Lyman and Werner band emissions excited by secondary electrons produced by cosmic rays.  Here we used the same methodology adopted in NWS05.
 
8) For the radiative recombination of Cl$^+$, we adopt the value recommended by Mazzotta et al.\ (1998).  Dielectronic recombination is negligible at the temperatures of present interest.

Our model does not include the photodissociation of molecular ions containing the halogen elements.  With the exception of HCl$^+$, for which photodissociation cross-sections have been presented by Pradhan, Kirby \& Dalgarno (1991) and by Andric et al.\ (2002), we are not aware of any estimates of the relevant cross-sections.  The HCl$^+$ photodissociation cross-sections imply an unshielded photodissociation rate of $5.6 \times 10^{-11} \chi_{UV} \rm s^{-1}$.  In the region where HCl$^+$ is most abundant, photodissociation makes a negligible contribution - relative to dissociative recombination - to the HCl$^+$ destruction rate; we assume photodissociation to be similarly negligible for CF$^+$ and H$_2$Cl$^+$.

The full reaction list for chlorine-bearing species is given in Table 1.  It contains 21 processes involving the following species: Cl, HCl, Cl$^+$, HCl$^+$, H$_2$Cl$^+$ and CCl$^+$.

\subsection{Chemical network for fluorine-bearing species}

The chemical network for fluorine-bearing species has been discussed by NWS05.  The chemistry is initiated by reaction of H$_2$ with F:
\begin{equation}
\rm F + H_2 \rightarrow HF + H
\end{equation}
Once the H$_2$ abundance becomes appreciable, HF becomes the dominant reservoir of fluorine; it is destroyed slowly by photodissociation, and by reactions with He$^+$, H$_3^+$ and C$^+$.  Reaction with C$^+$ leads to the formation of CF$^+$:
\begin{equation}
\rm HF + C^+ \rightarrow CF^+ + H
\end{equation}
The fluoromethylidynium ion, CF$^+$, is destroyed rapidly by dissociative recombination, but can nevertheless account for as much as $\sim 1\%$ of the gas-phase fluorine abundance.  It has recently been detected in the interstellar medium (Neufeld et al.\ 2006).

For completeness, we have implemented five updates to the chemical network presented by NWS05.  Of the following changes, the first three of which were noted in the UDFA reaction list (Woodall et al.\ 2007), only the first is expected to have any significant effect on the model predictions:

1) For the rate coefficient for dissociative recombination of CF$^+$, we now adopt the recent laboratory measurement of Novotny et al.\ (2005).  The adopted rate coefficient lies factors of 4 and 1.5 below that assumed by NWS05 at temperatures of 300 and 10 K respectively. 

2) We have corrected an erroneous value adopted by NWS05 for the rate coefficient for 
reaction of F with H$_2$O.  The correct value is $\rm 1.6 \times 10^{-11} \, cm^{-3} s^{-1}$.  This was stated correctly in the text of NWS05, but a value 10 times too large was erroneously given in the reaction list table (and erroneously implemented in our calculation).

3) For the reaction of H$_2$ with F$^+$, we include a channel that produces H$_2^+$ in addition to that which produces HF$^+$.

4) For the cosmic ray induced photodissociation of HF, we include the effects of H Lyman alpha radiation.  This results in a 50$\%$ increase in the adopted photodissociation rate.

5) For the radiative recombination and dielectronic recombination of F$^+$, we adopt the rate coefficients of Badnell (2006) and Badnell et al.\ (2003).

The full reaction list for fluorine-bearing species is given in Table 2.  It contains 18 processes involving the following species: F, HF, F$^+$, HF$^+$, H$_2$F$^+$, CF$^+$ and SiF$^+$.

\subsection{Interstellar cloud models}

As in NWS05, we have used a modified version of the PDR models of Kaufman et al.\ (2006), Kaufman et al.\ (1999), Wolfire et al.\ (1990), and  Tielens \& Hollenbach (1985) to investigate the chemistry within diffuse and dense molecular clouds.  The modifications to these models have been largely described in NWS05, with the exception of the following:

1) For the radiative and dielectronic recombination of C$^+$, we adopt the recently-computed rate coefficients of Badnell et al.\ (2006) and Badnell et al.\ (2003).

2) In treating the depletion of fluorine and chlorine that results from freeze-out onto grain mantles within dense molecular clouds, we use equation (5) of NWS05, but with the HF or HCl photodissociation rate augmented so as to include the effects of H$_2$ Lyman and Werner band emissions excited by the secondary electrons produced by cosmic rays. As in NWS05, we assume a gas-phase fluorine abundance of $1.8 \times 10^{-8}$ relative to H nuclei in diffuse molecular clouds.  For chlorine, we assume an analogous abundance of $1.8 \times 10^{-7}$, the average value observed in three sightlines that intersect diffuse molecular clouds: $\zeta$~Oph (Federman et al.\ 1995), HD 192639 (Sonnentrucker et al.\ 2002) and HD 185418 (Sonnentrucker, Friedman \& York 2006).

In our standard models, as in NWS05, we assume a {\it primary} cosmic-ray ionization rate of $\zeta_p = 1.8 \times 10^{-17}\,\rm s^{-1}$ per H {\it nucleus}.  Recent observations of H$_3^+$ have led to the postulate of substantially larger rates in diffuse clouds (e.g. Indriolo et al.\ 2007).  To investigate the effect of enhanced cosmic ray ionization rates upon the chemistry of interstellar halogens, we have also considered models for which the assumed $\zeta_p$ is increased by a factor 10. 

Our standard models also compute the thermal balance and equilibrium gas temperature as a function of distance into the cloud.  To investigate the possible effects of enhanced temperatures (such as might result from the passage of interstellar shocks), we have also considered a series of models in which the temperature is set to some depth-independent value (larger than that computed by our self-consistent treatment of the thermal balance). 

For our standard models, we have followed previous authors in adopting a branching ratio of 0.1 for the production of HCl following dissociative recombination of H$_2$Cl$^+$.  To study the dependence upon this branching ratio, 
we have also considered diffuse cloud models in which values of 0, 0.3 and 1.0 were adopted.

\section{Results}

We have constructed a series of one-sided models in which ultraviolet radiation is incident upon a semi-infinite slab, with the specific intensity constant over the incoming hemisphere.  Each model is characterized by two key parameters: the density of H nuclei, $n_{\rm H} = n({\rm H}) + 2\,n({\rm H_2})$, assumed constant throught the slab; and the incident radiation field, $\chi_{UV}$, expressed in units of the mean ISRF given by Draine (1978). 
For one-sided slab models, we have considered all combinations of $n_{\rm H} = 10^3$, $10^4$, $10^5$, $10^6$, and $10^7 \rm \, cm^{-3}$ and $\chi_{UV} = 10^2$, $10^3$, $10^4$, and $10^5$.

In Figure 3, we present example results for a single such combination: $n_{\rm H}= 10^4 \rm \, cm^{-3}$, and $\chi_{UV}= 10^4$.  Here, we plot the relative abundances of H$_2$, Cl, Cl$^+$, HCl, HCl$^+$, H$_2$Cl$^+$, HF, and CF$^+$ as a function of the depth into the cloud.  CCl$^+$ fails to appear on this plot because its abundance is too low.  Each abundance is normalized relative to the relevant {\it gas-phase} elemental abundance (i.e.\ hydrogen for the case of H, chlorine for the case of Cl-containing molecules, fluorine for the case of F-containing molecules).  The depth is measured in terms of the visual extinction, $A_V$, in magnitudes, along a perpendicular ray to the cloud surface.  Figure 4 shows the gas temperature and the F and Cl gas-phase elemental abundances as a function of depth.

In Figures 3 and 4, solid lines show the results for an assumed primary cosmic ray ionization rate of $1.8 \times 10^{-17} \,\rm s^{-1}$ per H nucleus (standard case), while dashed lines show the results for an assumed primary cosmic ray ionization rate of $1.8 \times 10^{-16}\,\rm s^{-1}$ per H nucleus.  The effect of the increased cosmic ray ionization rate is negligible for $A_V \le 3$, because the ionization of Cl is dominated by the ISRF near the cloud surface.  At larger depths below the cloud surface, the predicted abundances of $\rm H_2Cl^+$ and $\rm CF^+$ are significantly enhanced if a value of $1.8 \times 10^{-16}\,\rm s^{-1}$ is adopted for $\zeta_p$; however, at these depths, the observed H$_3^+$ abundances do {\it not} support an enhanced value for the cosmic ray ionization rate (Indriolo et al.\ 2007).

By integrating the molecular abundances into a depth $A_V = 10$~mag, we have obtained the total column densities for each case.  In Figure 5, we present the column densities of HCl, HCl$^+$, H$_2$Cl$^+$, HF, and CF$^+$.  Here, we have plotted the molecular column densities as a function of $n_{\rm H}$.  Results are shown for four different values of the radiation field, $\chi_{UV}= 10^2$, $10^3$, $10^4$, and $10^5$.  

We have also constructed two-sided models, in which a slab of finite thickness is illuminated from both sides.  While the one-sided models are primarily applicable to PDRs in which a hot star or star cluster irradiates a nearby molecular cloud, the two-sided models are applicable to diffuse or translucent molecular clouds that are exposed to the ISRF.  In Figure 6, we present the predicted column densities as a function of the total $A_{V,tot}$ across the slab.  Results are shown for clouds with density $n_{\rm H} = 10^{2.5} \, \rm \, cm^{-3}$ exposed to radiation with $\chi_{UV}$ = 1 and 10.  In Figure 7, we present the same model predictions but with $N({\rm H}_2)$ in place of $A_{V,tot}$ as the independent variable.

Figure 8 shows how the HCl and CCl$^+$ abundances are enhanced at elevated gas temperatures. Here we show the results of one-sided models for slabs of 
density $n_{\rm H} = 10^{2.5} \, \rm \, cm^{-3}$ exposed to radiation with $\chi_{UV}$ = 1 and 10.  In addition to the standard model in which the temperature is determined by considerations of thermal balance, we show the abundance profiles for models in which the temperature is arbitrarily set to 150, 200, 300 or 400 K.  Even temperatures as low as 150~K enhance the HCl abundance by driving the slightly endothermic reaction of Cl and H$_2$.  The CCl$^+$ abundance is increased accordingly, a result of the subsequent reaction of HCl with C$^+$.

\section{Discussion}

As expected, the results shown in Figures 3 -- 7 indicate qualitatively-different behavior for the F- and Cl-bearing species.  The results for F-bearing molecules are qualitatively the same as those obtained previously by NWS05: once H$_2$ becomes abundant, HF becomes the dominant reservoir of gas-phase fluorine, with CF$^+$ the only other significant molecule containing fluorine.  Because of the lower dissociative recombination rate assumed for CF$^+$ in the present work, the predicted CF$^+$ abundances are somewhat larger.  The chemistry for Cl-bearing molecules is somewhat more complex, by contrast, and the computer-generated network diagrams shown in Figure 9, together with Figures 10 - 13 in the online materials, help elucidate the key pathways.

\subsection{Identification of key chemical pathways}

In Figures 9 -- 13, circles represent the key Cl-bearing species, with a color scale indicating their relative abundance.  The color map extends from cyan to blue to purple to red (highest value), with grey circles denoting species that account for less than 10$^{-8}$ of the gas-phase chlorine abundance.
Arrows indicate the key reactions linking the various species: again, the colors of the arrows indicate the reaction rate per unit volume.
The results apply to a one-sided model with $n_{\rm H} = 10^4\,\rm cm^{-3}$ and $\chi_{UV} = 10^4$, and the different figures show the behavior at various depths into the cloud.  The depths were chosen to illustrate several different regimes.

1) $A_V = 0$:  At the cloud surface, there is no shielding and the abundance of molecules is small.  The Cl$^+$/Cl ratio is $\sim 10^5$, being determined by the balance between photoionization and recombination. For the latter process, radiative recombination and charge transfer recombination with H -- although endothermic by $\sim$~0.6 eV -- are of roughly equal importance. 

2) $A_V= 0.8$:  Here, H$_2$ self-shielding has greatly increased the H$_2$ abundance.  Reaction with H$_2$ is now the dominant destruction mechanism for Cl$^+$, resulting in the formation of HCl$^+$.  However, the e/H$_2$ abundance ratio is still sufficient for dissociative recombination to dominate the destruction of HCl$^+$, although the competing reaction with H$_2$ produces a small amount of H$_2$Cl$^+$.  The dissociative recombination of HCl$^+$ leads to atomic Cl, which becomes the dominant gas-phase reservoir of chlorine. 
Beyond $A_V \sim 0.8$, the abundances of the HCl$^+$ and H$_2$Cl$^+$ ions start to fall, along with that of $\rm Cl^+$.

3) $A_V = 2.0$: The e/H$_2$ abundance ratio has now dropped sufficiently that HCl$^+$ reacts primarily with H$_2$, rather than electrons, forming H$_2$Cl$^+$.  HCl is produced by dissociative recombination of H$_2$Cl$^+$, although the dominant formation process for HCl is the slightly endothermic reaction of H$_2$ and Cl; this reaction is still possible at the $\sim 200$~K temperature at this point within the PDR.  At $A_V = 2.0$, Cl and HCl account respectively for $\sim 0.1 \%$ and $\sim 99.9 \%$ of elemental chlorine in the gas phase.

4) $A_V = 4.0$: The gas temperature drops below the level at which HCl can be formed by reaction of H$_2$ with Cl, and the HCl abundance is lower than it was at $A_V = 2.0$.  The Cl$^+$/Cl ratio is now only $10^{-8}$, the photoionization rate of Cl having dropped dramatically.  Thus the formation of HCl$^+$ is dominated by reaction of H$_3^+$ with Cl, not H$_2$ with Cl$^+$.  The H$_3^+$ molecular ion is produced by the effects of cosmic rays; these ionize H$_2$, forming H$_2^+$, which then transfers a proton to H$_2$ to form H$_3^+$.  

5) $A_V = 10.0$:  As at $A_V=4.0$, HCl$^+$ is produced primarily by reaction of H$_3^+$ with Cl, and reaction of  HCl$^+$ with H$_2$ then forms H$_2$Cl$^+$.  At this depth, the electron fraction is so low that dissociative recombination of H$_2$Cl$^+$ is no longer dominant.  H$_2$Cl$^+$ transfers a proton to a neutral species of higher proton affinity than HCl (e.g. CO or H$_2$O; this process is denoted by the arrow labeled ''N''), yielding HCl.  At this point, the external UV is almost completely attenuated, and the HCl destruction rate is very small.  HCl now accounts for several $\times 10\%$ of the gas-phase chlorine abundance, with atomic chlorine accounting for essentially all of the remainder.  Our results in this regime are in good agreement with those obtained previously by Schilke et al\ (1995).
H$_2$Cl$^+$ and CCl$^+$ are also present, but their abundances are just a few $\times 0.01\%$ of gas-phase elemental chlorine.  HCl reacts with H$_3^+$ to form H$_2$Cl$^+$, although this rapidly leads to reformation of HCl by proton transfer to other neutral species.  Several other destruction processes are of comparable importance for HCl destruction:  cosmic ray induced photodissociation, and reaction with the positive ions C$^+$ and He$^+$.  In this regime, the rates of HCl formation and destruction all scale linearly with the cosmic-ray ionization rate, with the result that the HCl abundance is almost independent of the latter.

Although the diagrams shown in Figure 9 -- 13 apply to just a single model, similar regimes exist for other values of $\chi_{UV}$ and $n_{\rm H}$.  Although the relevant regions move in or out (to larger or smaller $A_V$) as the assumed $\chi_{UV} / n_{\rm H}$ ratio is increased or decreased, the same qualitative behavior is observed.  In examining similar diagrams for the entire set of models, we have not identified any important reaction pathway that is not apparent in Figures 9 -- 13.  

\subsection{Observational implications}

To date, HF, CF$^+$, and HCl are the only halogen-bearing molecules to have been detected in the interstellar medium.  In regard to HF, our results are identical to those presented in NWS05: we predict HF to be the dominant gas-phase reservoir of fluorine within both diffuse and dense molecular clouds; we expect the {\it Herschel Space Observatory} to detect widespread absorption in the HF $J=1-0$ transition. However, the abundances we predict for CF$^+$ lie a factor $\sim 3$ above the prediction of NWS05, a direct consequence of the smaller dissociation rate adopted for CF$^+$ in the present study.  As noted by Neufeld et al. (2006), the CF$^+$ column densities inferred from observations of the Orion Bar were already a factor of $\sim 4$ below the predictions of NWS05, and the discrepancy between theory and observation is now increased to more than an order of magnitude.  This disagreement may indicate that we have overestimated the rate coefficient for reaction of C$^+$ and HF. We are not aware of any laboratory studies of this reaction; the rate we adopt is simply the capture rate, and thus an upper limit on the true reaction rate.  Laboratory measurements of this key reaction would be very desirable.

HCl has been observed in both diffuse and dense molecular clouds.  In diffuse clouds, ultraviolet absorption studies have led to upper limits, and in one case, a tentative detection toward $\zeta$ Oph (Federman et al.\ 1995).  The latter result, obtained using the Goddard High Resolution Spectrograph on HST, yielded an HCl column density of $2.7 \pm 1.0 (1\sigma) \times 10^{11} \rm cm^{-2}$.  Given an atomic chlorine column density of $3.0 \pm 1.0 (1\sigma) \times 10^{14} \rm cm^{-2}$ for this sight-line (Federman et al.\ 1995), the corresponding $N({\rm HCl})/N({\rm Cl})$ ratio is $9 \times 10^{-4}$ (uncertain by a factor $\sim 2$), a value in excellent agreement with predictions of the diffuse cloud models presented by vDB86.   Very similar results are obtained from our updated treatment of Cl chemistry.  In Figure 14, we show predictions from a series of models with $\chi_{UV}$ ranging from $10^{-1}$ to 10$^2$.  All models apply to a slab with parameters appropriate to $\zeta$~Oph, viz.\ density $n_H = 10^{2.5} \rm \, cm^{-3}$ and total visual extinction of 0.8 mag.  The horizontal axis shows the H$_2$ column density, for which the measured value is $4.2 \times 10^{20}$ along this sight-line, while the vertical axis shows $N({\rm HCl})/N({\rm Cl})$ (red curve), $N({\rm Cl^+})/N({\rm Cl})$ (orange) and $N({\rm HCl^+})/N({\rm Cl})$ (blue).  Squares located along the curves show the results for $\chi_{UV} = 10^{-1}$ to 10$^2$ from right to left.  The H$_2$ column density along the $\zeta$~Oph sight-line requires a $\chi_{UV} \sim 10^{0.5}$, in agreement with previous studies, and the predicted $N({\rm HCl})/N({\rm Cl})$ is in good agreement with the observations.  The almost exact agreement between the present model and that of vDB86 in regard to the predicted HCl abundance appears to result from the fortuitious cancellation of two changes in the photochemical network: the photodissociation rate for HCl and the photoionization rate for Cl are both increased by a factor $\sim 2$.
As in the vDB86 models, the predicted Cl$^+$ column density lies substantially below the observations.   Federman et al.\ (1995) attributed this discrepancy to the presence of a significant Cl$^+$ column density within  HII regions along the sight-line. 
One important caveat must be noted.  The results shown in Figure 14 were obtained by assuming a branching ratio of only 10$\%$ for the production of HCl following dissociative recombination of H$_2$Cl$^+$.  This value was initially invoked primarily to {\it explain} the relative low HCl abundance derived from upper limits obtained with the {\it Copernicus} satellite.  Models that we obtained for other branching ratios (0, 0.3 and 1.0) indicate that the HCl column density is linearly proportional to the branching ratio for values greater than 0.1  With a branching ratio of zero (i.e.\ with no production of HCl from dissociative recombination of H$_2$Cl$^+$), a non-zero but negligible HCl column density ($N({\rm HCl})/N({\rm Cl}) \sim 10^{-6}$) results from the endothermic reaction of H$_2$ with Cl.

In dense clouds, HCl has been unequivocally detected in several sources by means of submillimeter observations of the $J=1-0$ emission line.  First detected using the Kuiper Airborne Observatory toward Orion (Blake, Keene \& Phillips 1985) and then Sgr B2 (Zmuidzinas et al.\ 1995), the $J=1-0$ transition has also been observed using the ground-based Caltech Submillimeter Observatory (CSO) under good atmospheric conditions.  CSO observations have led to additional detections toward Orion A, Mon R2 (Salez, Frerking \& Langer 1996) and several other sources (Phillips et al.\ 2009), as well as mapping observations toward OMC-1 (Schilke, Phillips \& Wang 1995).  Typical column densities derived for these dense clouds lie in the few $\rm \times 10^{13}$ to few $\times 10^{14} \, \rm cm^{-2}$ range, in reasonable agreement with the predictions of our model for regions at high density exposed to strong UV radiation (Fig.\ 5).  As noted in the observational studies cited above, depletion plays an important role in limiting the fractional abundance of HCl.

Our study identifies two additional Cl-bearing species that are potentially detectable: the molecular ions HCl$^+$ and H$_2$Cl$^+$.   These ions are isoelectronic with OH and H$_2$S respectively.  They are most abundant near cloud surfaces, where the photoionization rate for HCl is highest, and show column densities that are an increasing function of $\chi_{UV} / n_{\rm H}$.  Thus, PDRs subject to strong UV irradiation present attractive targets for searches for HCl$^+$ and H$_2$Cl$^+$ at millimeter and submillimeter wavelengths.  For example, in a PDR with $n_{\rm H} = 10^4\,\rm cm^{-3}$ and $\chi_{UV} = 10^4$, column densities of $5.7$ and $2.6\times 10^{11} \, \rm cm^{-2}$ are predicted for HCl$^+$ and H$_2$Cl$^+$.  These values apply to sight-lines perpendicular to the illuminated surface; in edge-on PDRs, they can be significantly enhanced by limb-brightening.  The rotational spectrum of HCl$^+$ has been measured using laser magnetic spectroscopy (Lubic et al.\ 1989).  The lowest-lying rotational transition, the $^2\Pi_{3/2} J = 5/2 \rightarrow 3/2$ multiplet near 207.6$\mu$m, lies in a wavelength range accessible to the HIFI instrument on the {\it Herschel Space Observatory}.  The rotational spectrum of H$_2$Cl$^+$ has also been the subject of a laboratory study at high spectral resolution; here, the measurements of Araki et al.\ (2001) indicate that the lowest-lying rotational line of ortho-H$_2$Cl$^+$, the $1_{10}-1_{01}$ transition, lie near 189.2 and 188.4 GHz respectively for the  $\rm H_2^{35}Cl^+$ and $\rm H_2^{37}Cl^+$ isotopologues.  (These transitions are split into 6 and 4 hyperfine components, covering $\sim 56$ and $14$~MHz respectively.)  This spectral region lies in the wing of the strong 183 GHz telluric water absorption line, but is nevertheless observable from ground-based observatories under favorable -- but by no means unusual -- atmospheric conditions.   In one of the  earliest studies of interstellar chlorine chemistry, the possibility of detecting HCl$^+$ in {\it diffuse} molecular clouds was discussed by Jura (1974), who considered the $A^2\Sigma^+ \leftarrow X ^2\Pi$ band that is accessible in the near-ultraviolet region.   Given an oscillator strength of $4.15 \times 10^{-4}$ for the strongest vibrational band (Pradhan, Kirby \& Dalgarno 1991), and given the abundances predicted in Figure 17, our results for $\zeta$~Oph would imply equivalent widths of only $\sim 4 \times 10^{-6}\,\AA$, well below the limit of detectability.  

Finally, we note that, unlike CF$^+$, the CCl$^+$ ion is typically predicted to have an extremely low abundance.  Although CCl$^+$ is produced rapidly by reaction of HCl and C$^+$, these two species show very little overlap: the HCl abundance is small in the surface layers where C$^+$ is abundant, unless the temperature is increased by shock heating, and the C$^+$ abundance is small in the deep cloud interiors where HCl is relatively abundant.  Enhanced CCl$^+$ abundances are possible when shock heating is present along with UV irradiation.

\begin{acknowledgments}

We are very grateful to W.~D.~Geppert and M.~Hamberg for informing us about recent laboratory results on H$_2$Cl$^+$ dissociative recombination, in advance of their publication.  We thank Paule Sonnentrucker and Paul Bryans for several useful discussions. 
We gratefully acknowledge the support of a grant from the NASA Herschel Science Center's Theoretical Research/Laboratory Astrophysics Program.

\end{acknowledgments}

\begin{deluxetable}{ll}
\rotate
\tablewidth{0pt}
\tablecaption{Reaction list for Cl-bearing species}
\tablehead{\colhead{Reaction} & Rate or rate coefficient \\}
\startdata
$\rm Cl + H_2 \rightarrow HCl + H  \phantom{00}$ & $2.52 \times \rm 10^{-11}\,\exp\,(-2214\,K/{\it T})\, cm^3 \,s^{-1}$ 
$\phantom{0000000000000} \rm ({\it T} \le 354\, K)$\\
& $\rm 2.86 \times \rm 10^{-12}\,({\it T}/300\,K)^{1.72}\,\exp\,(-1544\,K/{\it T})\, cm^3 \,s^{-1}$ 
$\,\, \rm ({\it T} \ge 354 \,K)$\\
$\rm HCl + H_2 \rightarrow Cl + H_2  \phantom{00}$ & $1.49 \times \rm 10^{-11}\,\exp\,(-1763\,K/{\it T})\, cm^3 \,s^{-1}$ 
$\phantom{0000000000000} \rm ({\it T} \le 354\, K)$\\
& $\rm 1.69 \times \rm 10^{-12}\,({\it T}/300\,K)^{1.72}\,\exp\,(-1093\,K/{\it T})\, cm^3 \,s^{-1}$ 
$\,\, \rm ({\it T} \ge 354 \,K)$\\
$\rm C^+ + HCl \rightarrow CCl^+ + H  $ & $1.84 \times 10^{-9} \, (T/300 \rm \, K)^{-0.202}
\, cm^3 \,s^{-1}$\\
$\rm H_3^+ + HCl \rightarrow H_2 + H_2Cl^+ $ & $6.35 \times 10^{-9} \, (T/300 \rm \, K)^{-0.202}
\, cm^3 \,s^{-1}$\\
$\rm He^+ + HCl \rightarrow H + Cl^+ + He \phantom{000000}$ & $5.51 \times 10^{-9} \, (T/300 \rm \, K)^{-0.202}
\, cm^3 \,s^{-1}$\\
$\rm Cl^+ + H_2 \rightarrow HCl^+ + H $ & $1.0 \times \rm 10^{-9} 
\, cm^3 \,s^{-1}$\\
$\rm HCl^+ + H_2 \rightarrow H_2Cl^+ + H $ & $1.3 \times \rm 10^{-9} 
\, cm^3 \,s^{-1}$\\
$\rm CCl^+ + e \rightarrow C + Cl $ & $2.0 \times 10^{-7} \, (T/300 \rm \, K)^{-0.5}
\, cm^3 \,s^{-1}$\\
$\rm HCl^+ + e \rightarrow H + Cl $ & $2.0 \times 10^{-7}\, (T/300 \rm \, K)^{-0.5}
\, cm^3 \,s^{-1}$\\
$\rm H_2Cl^+ + e \rightarrow HCl + H $ & $1.2 \times 10^{-8}\, (T/300 \rm\, K)^{-0.85}
\, cm^3 \,s^{-1}$\\
$\rm H_2Cl^+ + e \rightarrow Cl + products$ & $1.08 \times 10^{-7}\, (T/300 \rm\, K)^{-0.85}
\, cm^3 \,s^{-1}$\\
$\rm H_2Cl^+ + CO \rightarrow HCl + HCO^+$ & $\rm 7.8 \times 10^{-10}\,
\, cm^3 \,s^{-1}$\\
$\rm H_2Cl^+ + H_2O \rightarrow HCl + H_3O^+$ & $3.7 \times 10^{-9}\, (T/300 \rm\, K)^{-0.194}
\, cm^3 \,s^{-1}$\\
$\rm Cl + H_3^+ \rightarrow H_2Cl^+ + H$ & $1.0 \times \rm 10^{-9} 
\, cm^3 \,s^{-1}$\\
$\rm HCl + h\nu \rightarrow H + Cl $ \phantom{0000000} & $\,1.7 \times 10^{-9}\,\chi_{UV} [\,{1 \over 2}E_2(2.1 A_V) + {1 \over 2}E_2(2.1 [A_{V,{\rm tot}}-A_V])]\,{\rm s}^{-1}\, $ \\
& \phantom{0000000}+ $\, 1370 \, \zeta_{p}\, /(1 - \omega) $ \\
$\rm HCl + h\nu \rightarrow HCl^+ + e$ & $\,1.0 \times 10^{-10}\,\chi_{UV} [\,{1 \over 2}E_2(3.45 A_V) + {1 \over 2}E_2(3.45 [A_{V,{\rm tot}}-A_V])]\,{\rm s}^{-1}\, $ \\
& \phantom{0000000}$ \times \, \exp(-N({\rm H}_2)/3.5 \times 10^{21} {\rm cm}^{-2}) /
(1 + [N({\rm H}_2)/3.2 \times 10^{20} {\rm cm}^{-2}]^{1/2})$ \\
& \phantom{0000000}+ $\, 185 \, \zeta_{p}\, /(1 - \omega) $ \\
\\
$\rm Cl + h\nu \rightarrow Cl^+ + e$ & $\,4.8 \times 10^{-11}\,\chi_{UV} [\,{1 \over 2}E_2(3.6 A_V) + {1 \over 2}E_2(3.6 [A_{V,{\rm tot}}-A_V])]\,{\rm s}^{-1}\, $ \\
& \phantom{0000000} $ \times \, \exp(-N({\rm H}_2)/4.28 \times 10^{21} {\rm cm}^{-2}) /
(1 + [N({\rm H}_2)/4.9 \times 10^{20} {\rm cm}^{-2}]^{1/2})$ \\
& \phantom{0000000} + $\, 61 \, \zeta_{p}\, /(1 - \omega) $ \\
$\rm Cl^+ + e \rightarrow Cl + h\nu $ & $1.34 \times 10^{-11}\, (T/300 \rm \, K)^{-0.738}
\, cm^3 \,s^{-1}$\\ 
$\rm Cl^+ + H \rightarrow Cl + H^+$   & $6.2 \times \rm 10^{-11}\, (T/300 \rm \, K)^{0.79}\,\exp\,(-6920\,K/{\it T})\, cm^3 \,s^{-1}$ \\
$\rm Cl + H^+ \rightarrow Cl^+ + H$   & $9.3 \times \rm 10^{-11}\, (T/300 \rm \, K)^{0.73}\,\exp\,(-232\,K/{\it T})\, cm^3 \,s^{-1}$ \\
$\rm HCl + H^+ \rightarrow HCl^+ + H$ & $3.3 \times \rm 10^{-9}\, (T/300 \rm \, K)^{1.00}\, cm^3 \,s^{-1}$ \\

\enddata
\end{deluxetable}
\clearpage
\begin{deluxetable}{ll}
\tablewidth{0pt}
\tablecaption{Reaction list for F-bearing species}
\tablehead{\colhead{Reaction} & Rate or rate coefficient \\}
\startdata
$\rm F + H_2 \rightarrow HF + H  \phantom{00}$ & $1.0 \times \rm 10^{-10}\,[\,\exp\,(-450\,K/{\it T})+ 0.078\,exp\,(-80\,K/{\it T})$ \\
& $\phantom{1.4 \times \rm 10^{-10}\,} \rm + 0.0155\,exp\,(-10\,K/{\it T})] 
\, cm^3 \,s^{-1}$\\
$\bf F + H_2O \rightarrow HF + O \, ^a$ & $\bf 1.6 \times 10^{-11} 
\, cm^3 \,s^{-1}$\\
$\rm C^+ + HF \rightarrow CF^+ + H  $ & $7.2 \times 10^{-9} \, (T/300 \rm \, K)^{-0.15}
\, cm^3 \,s^{-1}$\\
$\rm Si^+ + HF \rightarrow SiF^+ + H $ & $5.7 \times 10^{-9} \, (T/300 \rm \, K)^{-0.15}
\, cm^3 \,s^{-1}$\\
$\rm H_3^+ + HF \rightarrow H_2 + H_2F^+ $ & $1.2 \times 10^{-8} \, (T/300 \rm \, K)^{-0.15}
\, cm^3 \,s^{-1}$\\
$\rm He^+ + HF \rightarrow H + F^+ + He \phantom{000000}$ & $1.1 \times 10^{-8} \, (T/300 \rm \, K)^{-0.15}
\, cm^3 \,s^{-1}$\\
$\bf F^+ + H_2 \rightarrow H_2^+ + F $ & $\bf 6.24 \times \bf 10^{-10} 
\, cm^3 \,s^{-1}$\\
$\phantom{\bf F^+ + H_2} \bf \rightarrow HF^+ + H $ & $\bf 3.8 \times 10^{-10} 
\, cm^3 \,s^{-1}$\\
$\rm HF^+ + H_2 \rightarrow H_2F^+ + H $ & $1.3 \times \rm 10^{-9} 
\, cm^3 \,s^{-1}$\\
$\bf CF^+ + e \rightarrow C + F $ & $\bf 5.2 \times 10^{-8} \, (T/300 \bf \, K)^{-0.8}
\, cm^3 \,s^{-1}$\\
$\rm SiF^+ + e \rightarrow Si + F $ & $2.0 \times 10^{-7}\, (T/300 \rm \, K)^{-0.5}
\, cm^3 \,s^{-1}$\\
$\rm HF^+ + e \rightarrow H + F $ & $2.0 \times 10^{-7}\, (T/300 \rm \, K)^{-0.5}
\, cm^3 \,s^{-1}$\\
$\rm H_2F^+ + e \rightarrow HF + H $ & $3.5 \times 10^{-7}\, (T/300 \rm\, K)^{-0.5}
\, cm^3 \,s^{-1}$\\
$\rm H_2F^+ + e \rightarrow F + products$ & $3.5 \times 10^{-7}\, (T/300 \rm\, K)^{-0.5}
\, cm^3 \,s^{-1}$\\
$\rm F + CH \rightarrow HF + C$ & $1.6 \times \rm 10^{-10} 
\, cm^3 \,s^{-1}$\\
$\rm F + OH \rightarrow HF + O$ & $1.6 \times \rm 10^{-10}
\, cm^3 \,s^{-1}$\\
$\rm F + H_3^+ \rightarrow H_2F^+ + H$ & $4.8 \times \rm 10^{-10} 
\, cm^3 \,s^{-1}$\\
$\bf HF + h\nu \rightarrow H + F $ \phantom{0000000} & $\,1.17 \times 10^{-10}\,\chi_{UV} [\,{1 \over 2}E_2(2.21 A_V) + {1 \over 2}E_2(2.21 [A_{V,{\rm tot}}-A_V])]\,{\rm s}^{-1}\, $ \\
& + $\bf \, 124 \, \zeta_{p}\, /(1 - \omega) $ \\
\enddata
\tablenotetext{a}{Bold face denotes those reactions for which the adopted rates are different from those assumed by NWS05}
\end{deluxetable}
\clearpage

\begin{figure}
\includegraphics[scale=0.55,angle=0]{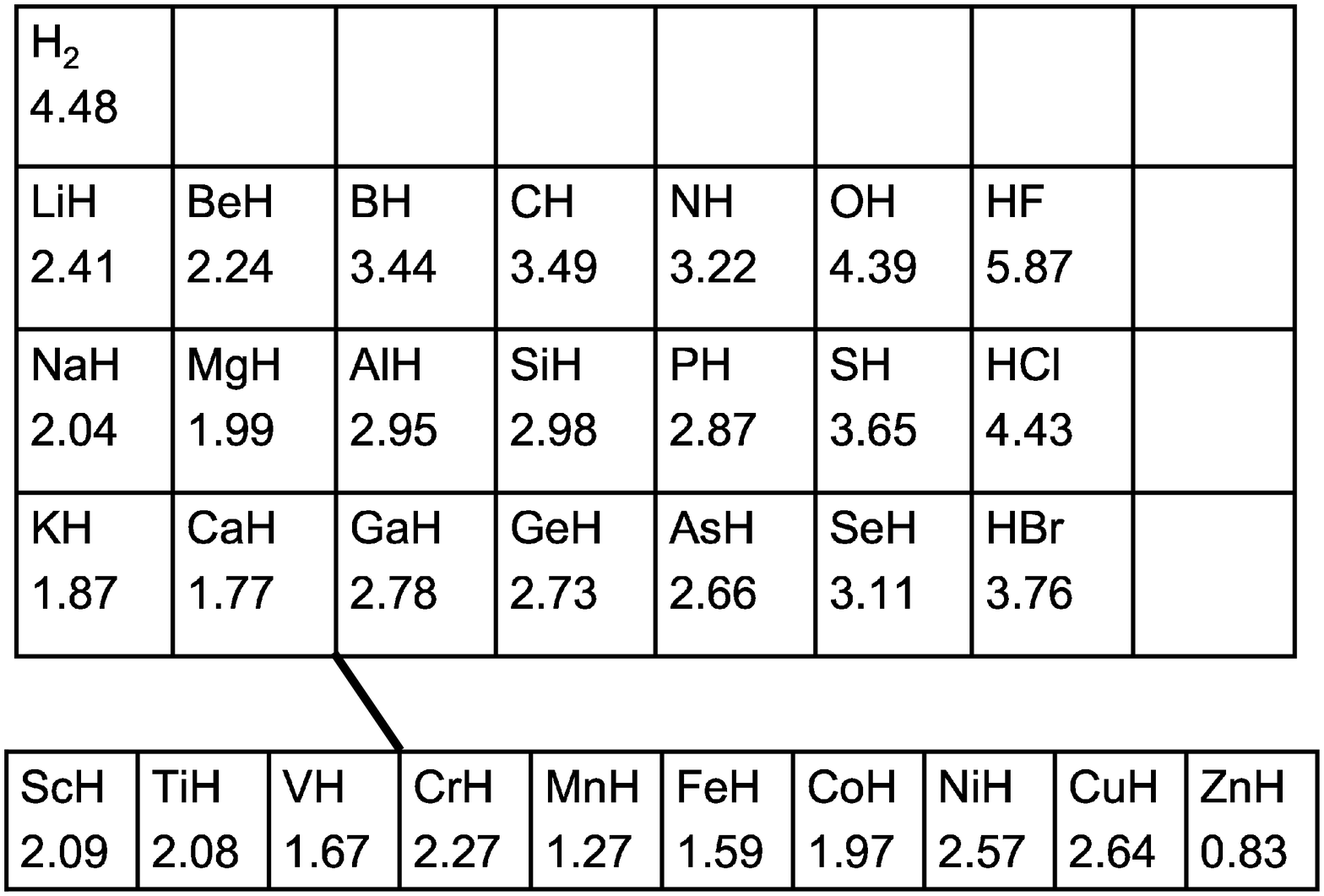}

\noindent{Fig.\ 1 -- Dissociation energies, $D^0_0$ (at 0~K in eV), for the neutral diatomic hydrides (derived from values recommended in the NIST Chemistry Web book and in Armentrout \& Sunderlin 1992, Chen, Clemmer \& Armentrout 1991, 1993; Binning \& Curtiss 1990;
and Huber \& Herzberg 1979)}
\end{figure}
\begin{figure}
\includegraphics[scale=0.9,angle=0]{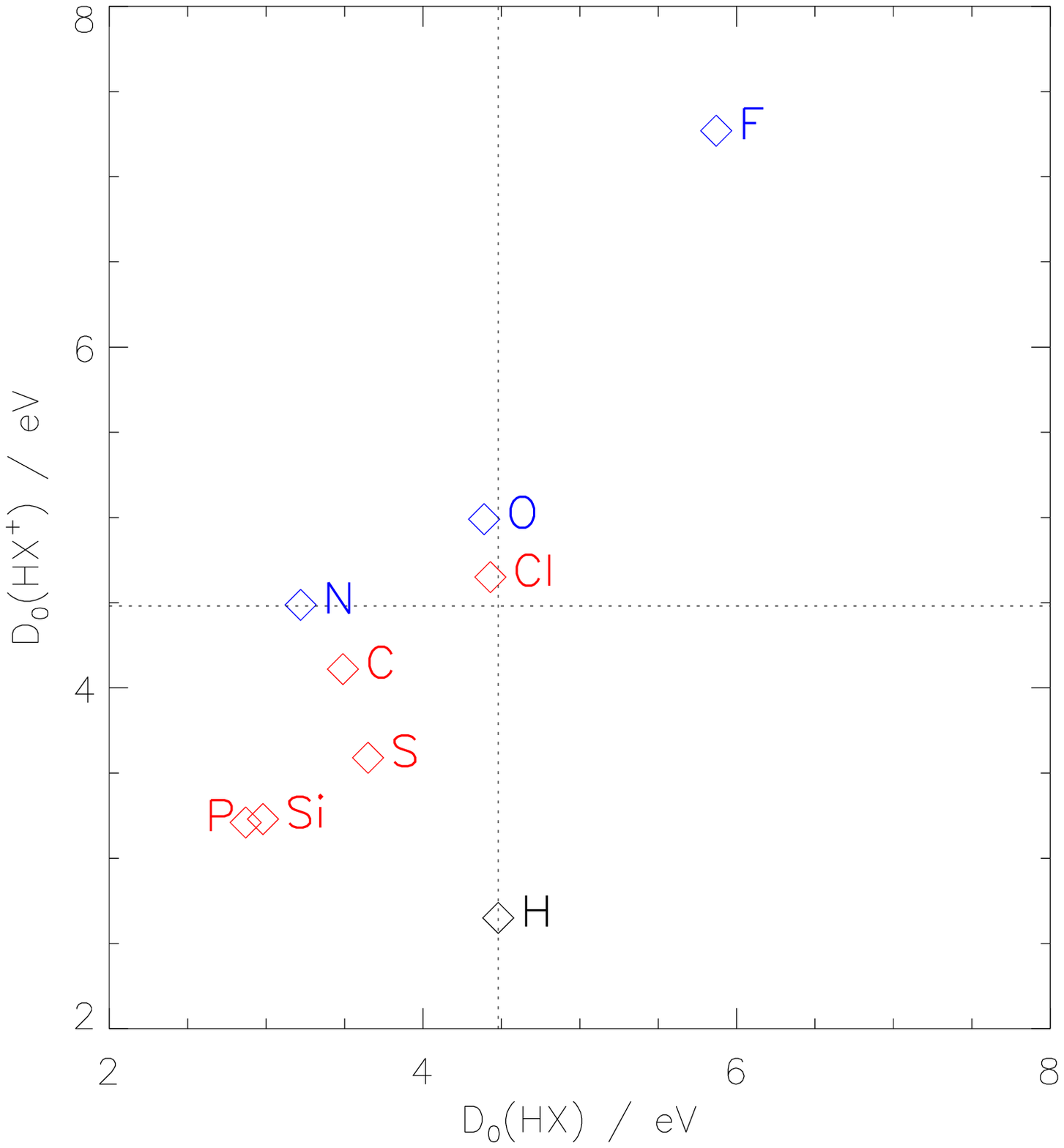}
\noindent{Fig.\ 2 -- Dissociation energies of HX and HX$^+$ for X = H, C, N, O, F, Si, P, S, and Cl.
Atoms with IP $<$ IP(H) appear in blue; those with IP $>$ IP(H) appear in red.  The dashed lines indicate
the dissociation energy of H$_2$ (4.48eV).}
\end{figure}
\begin{figure}
\includegraphics[scale=0.9,angle=0]{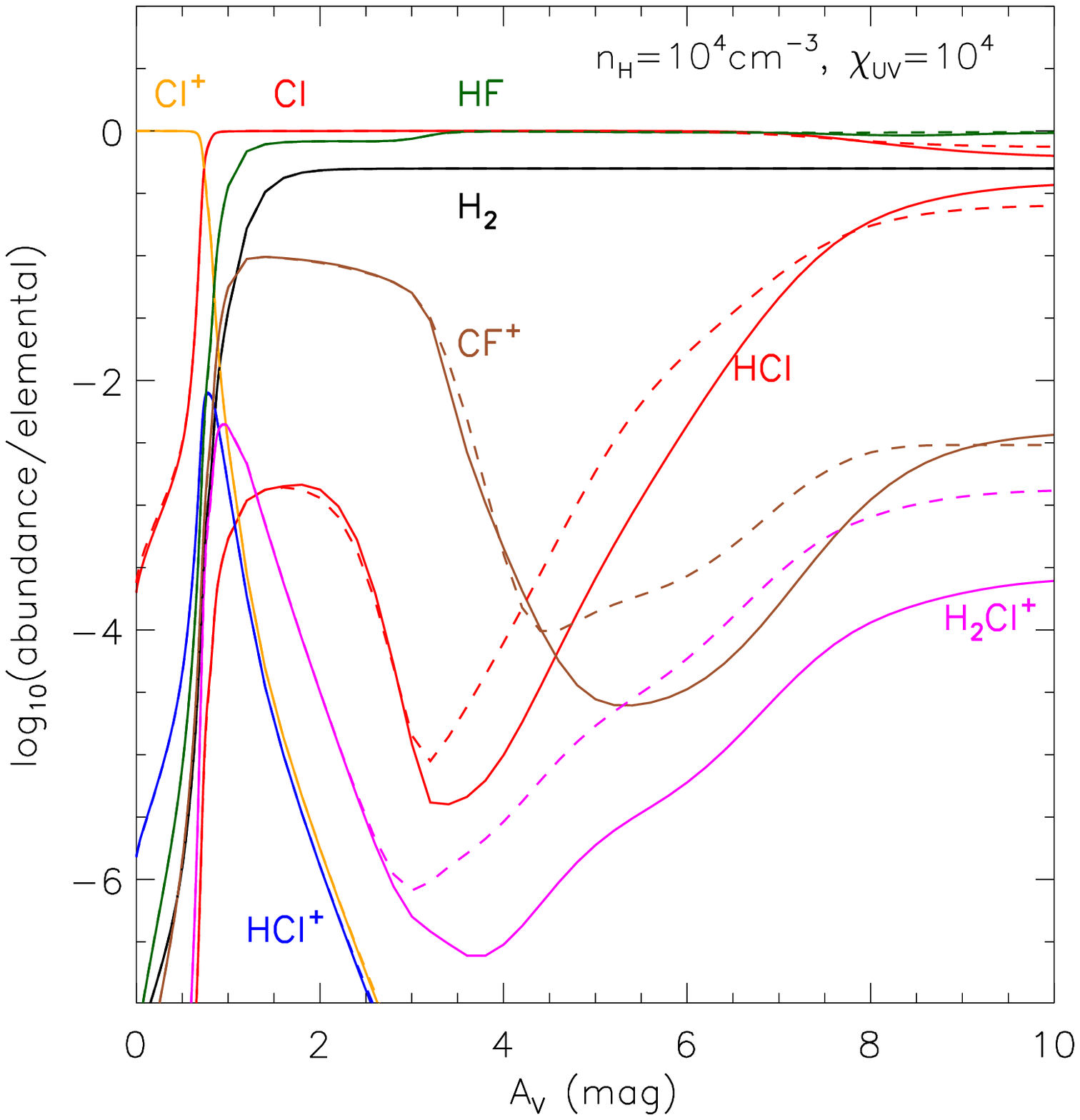}
\noindent{Fig.\ 3 -- Abundance profiles for a one-sided slab model with $\rm n_H=10^4\, cm^{-3}$ and $\chi_{UV}=10^4$.  Solid lines show the results for an assumed primary cosmic ray ionization rate of $1.8 \times 10^{-17}\,\rm s^{-1}$ per H nucleus (standard case).  Dashed lines show the results for an assumed primary cosmic ray ionization rate of $1.8 \times 10^{-16}\,\rm s^{-1}$ per H nucleus.}
\end{figure}
\begin{figure}
\includegraphics[scale=0.9,angle=0]{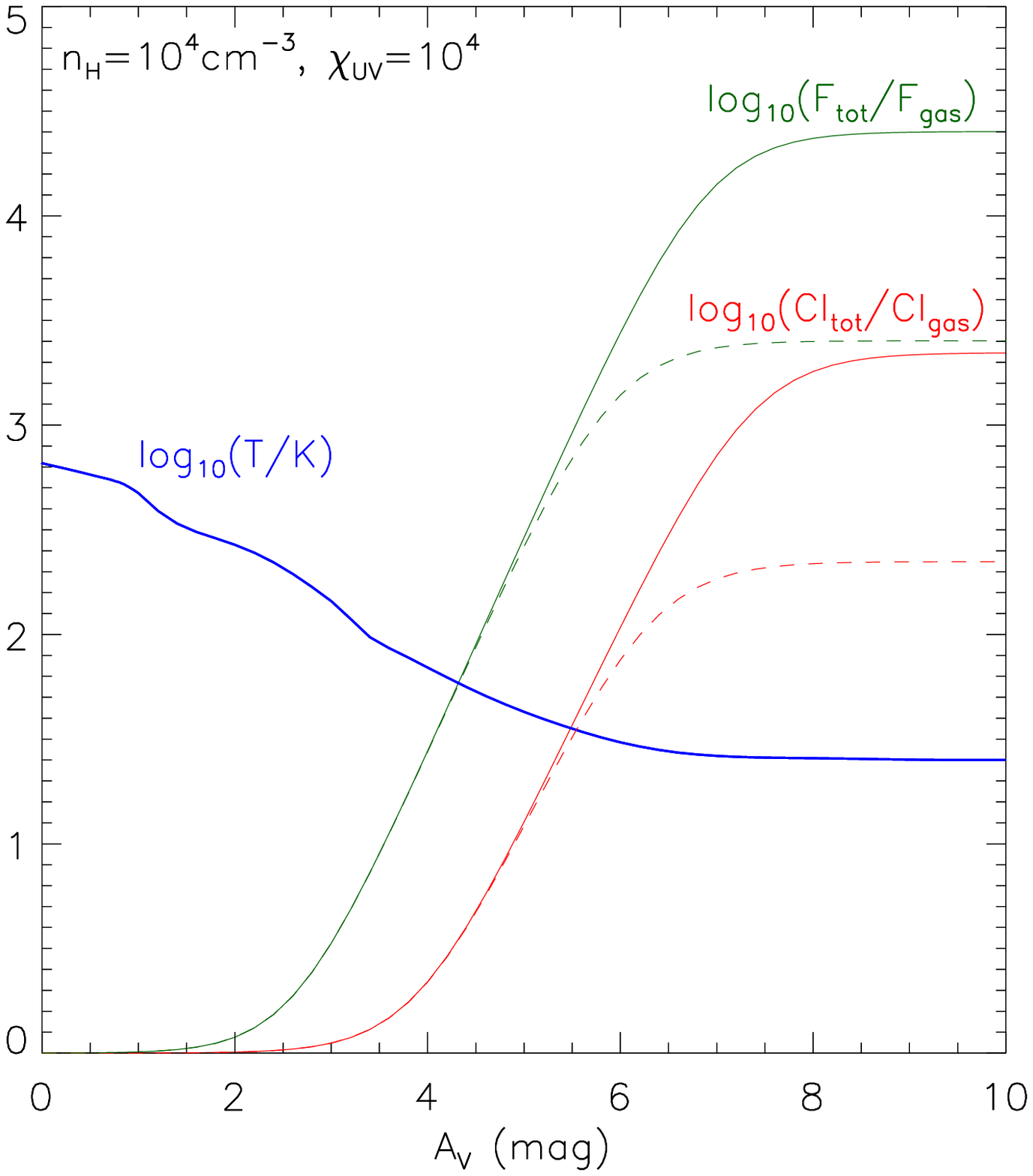}
\noindent{Fig.\ 4 -- Profiles of temperature and F- and Cl-depletion factor, for a one-sided slab model with $\rm n_H=10^4\, cm^{-3}$ and $\chi_{UV}=10^4$.  Solid lines show the results for an assumed primary cosmic ray ionization rate of $1.8 \times 10^{-17}\,\rm s^{-1}$ per H nucleus (standard case).  Dashed lines show the results for an assumed primary cosmic ray ionization rate of $1.8 \times 10^{-16}\,\rm s^{-1}$ per H nucleus.}
\end{figure}
\begin{figure}
\includegraphics[scale=0.9,angle=0]{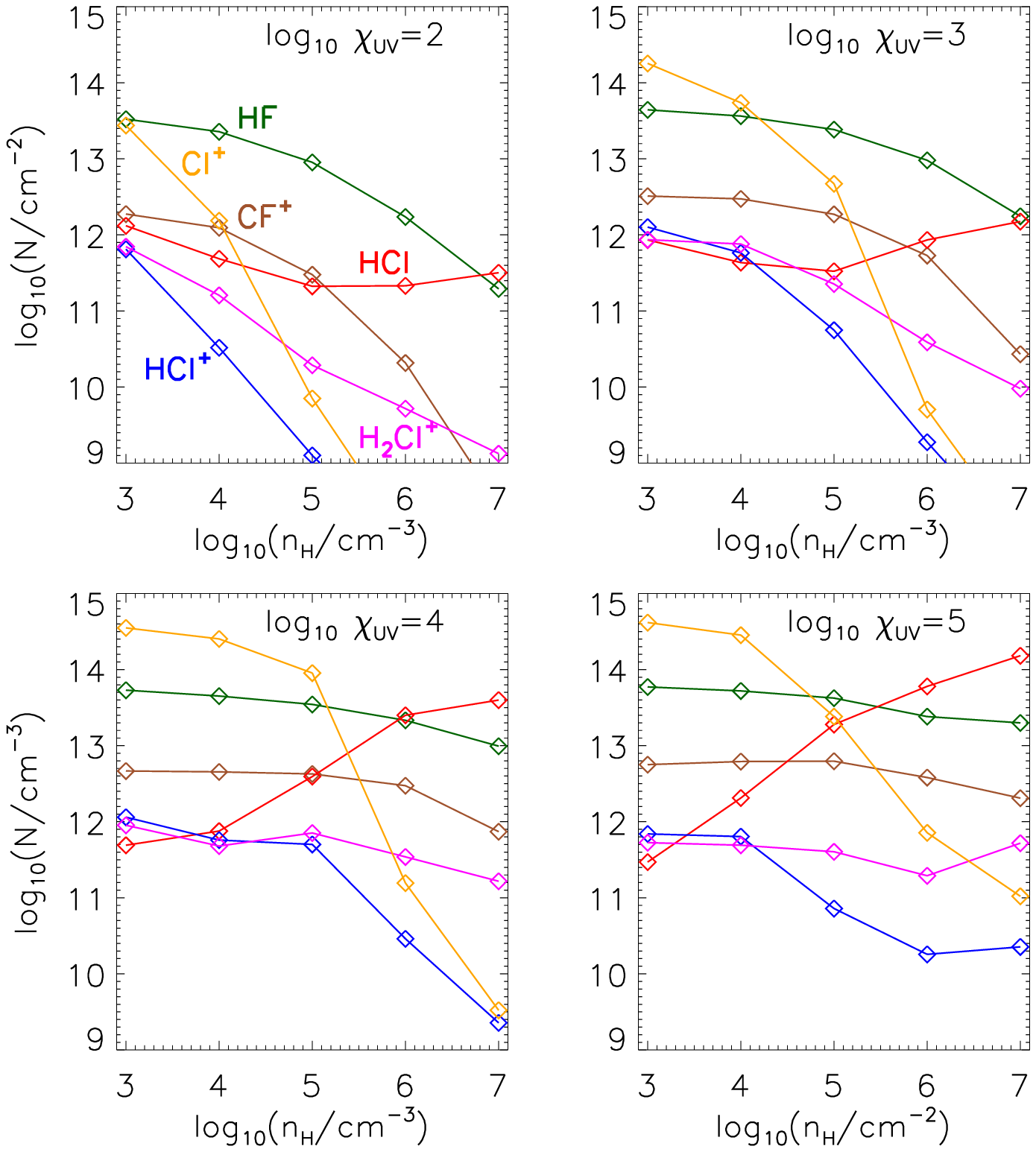}
\noindent{Fig.\ 5 -- Molecular column densities, for one-sided slab models}
\end{figure}
\begin{figure}
\includegraphics[scale=0.9,angle=0]{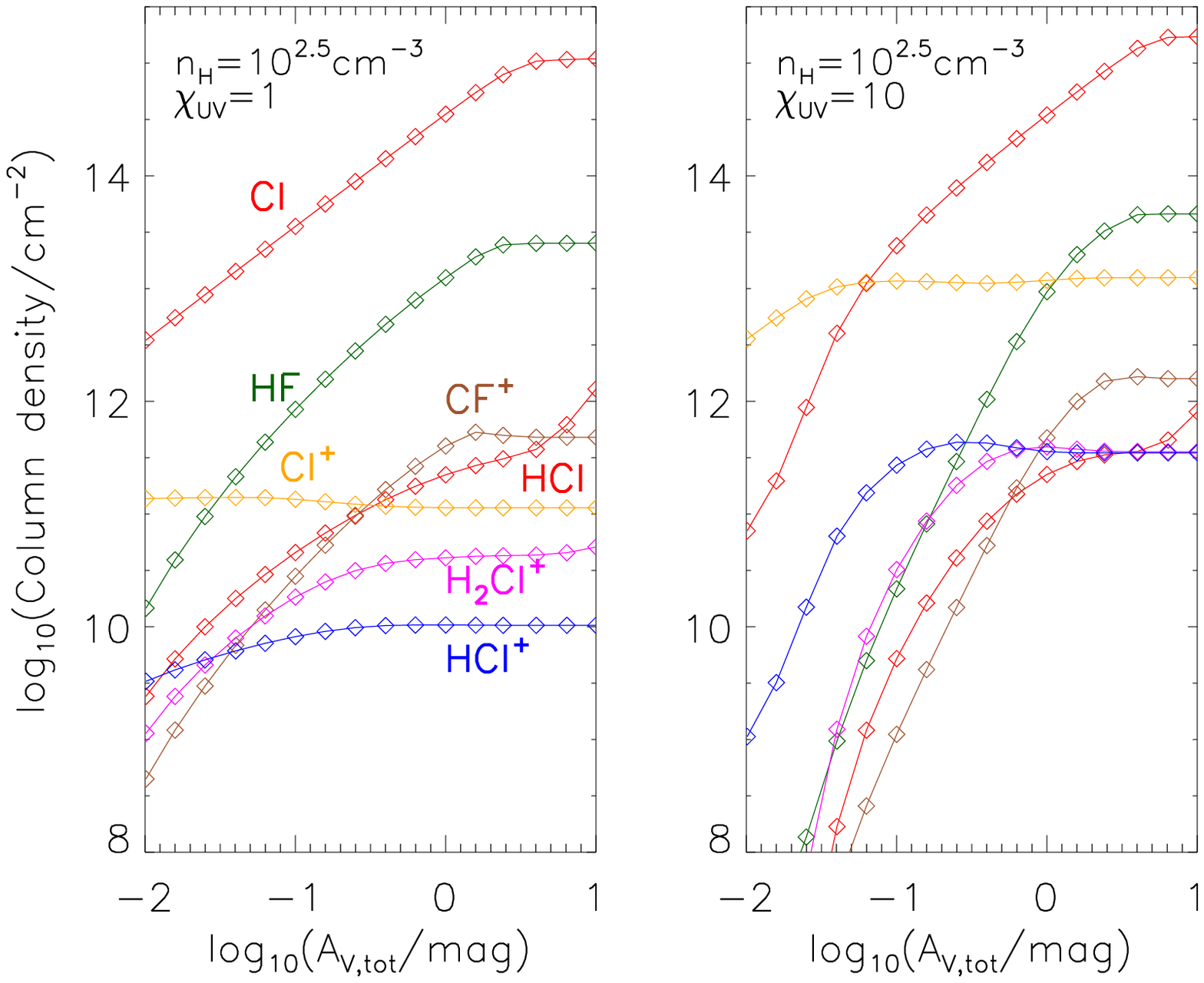}
\noindent{Fig.\ 6 -- Molecular column densities, for two-sided slab models, as a function of A$_V$ across the slab}
\end{figure}
\begin{figure}
\includegraphics[scale=0.9,angle=0]{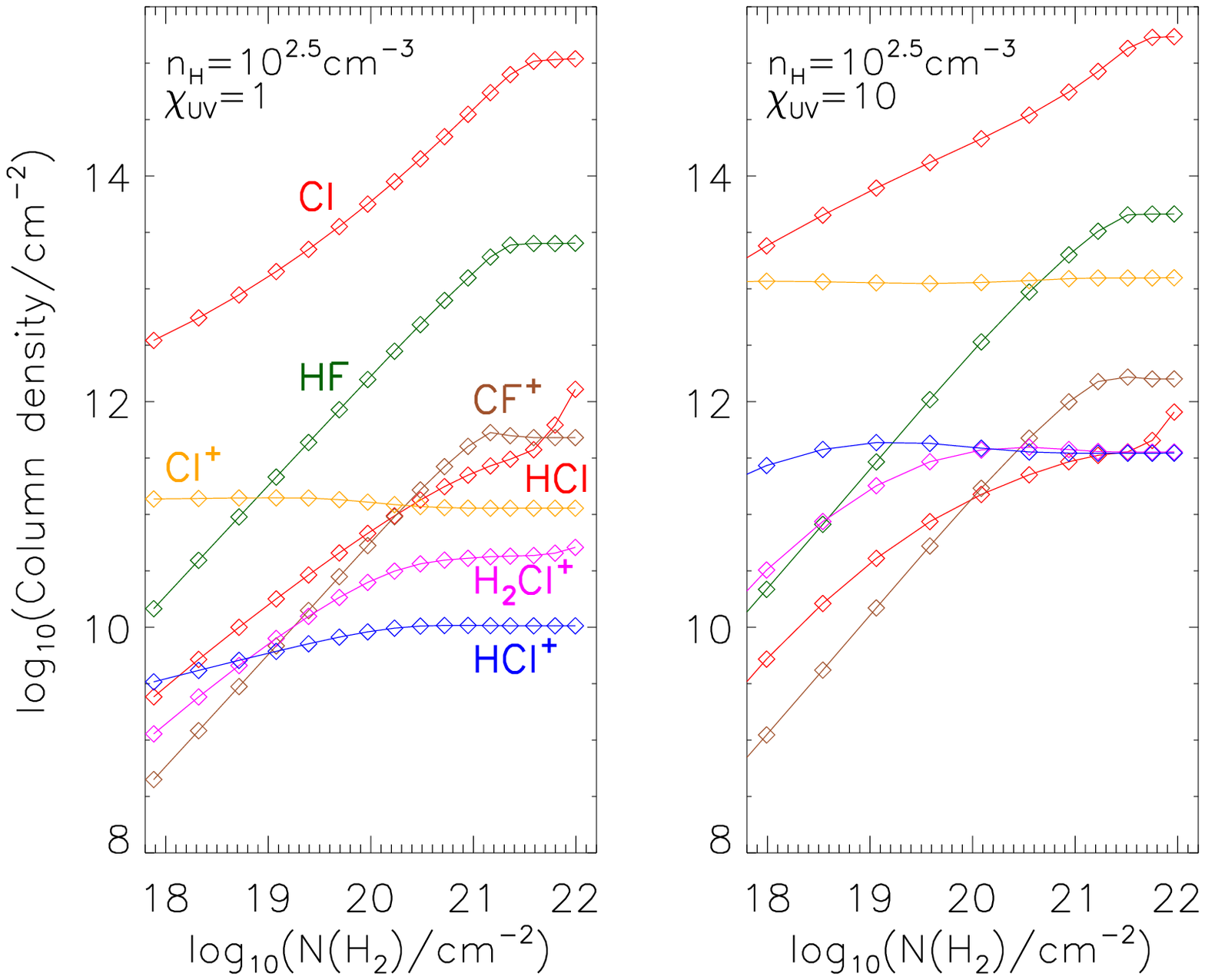}
\noindent{Fig.\ 7 --  Molecular column densities, for two-sided slab models, as a function of $N({\rm H}_2)$ across the slab}
\end{figure}
\begin{figure}
\includegraphics[scale=0.9,angle=0]{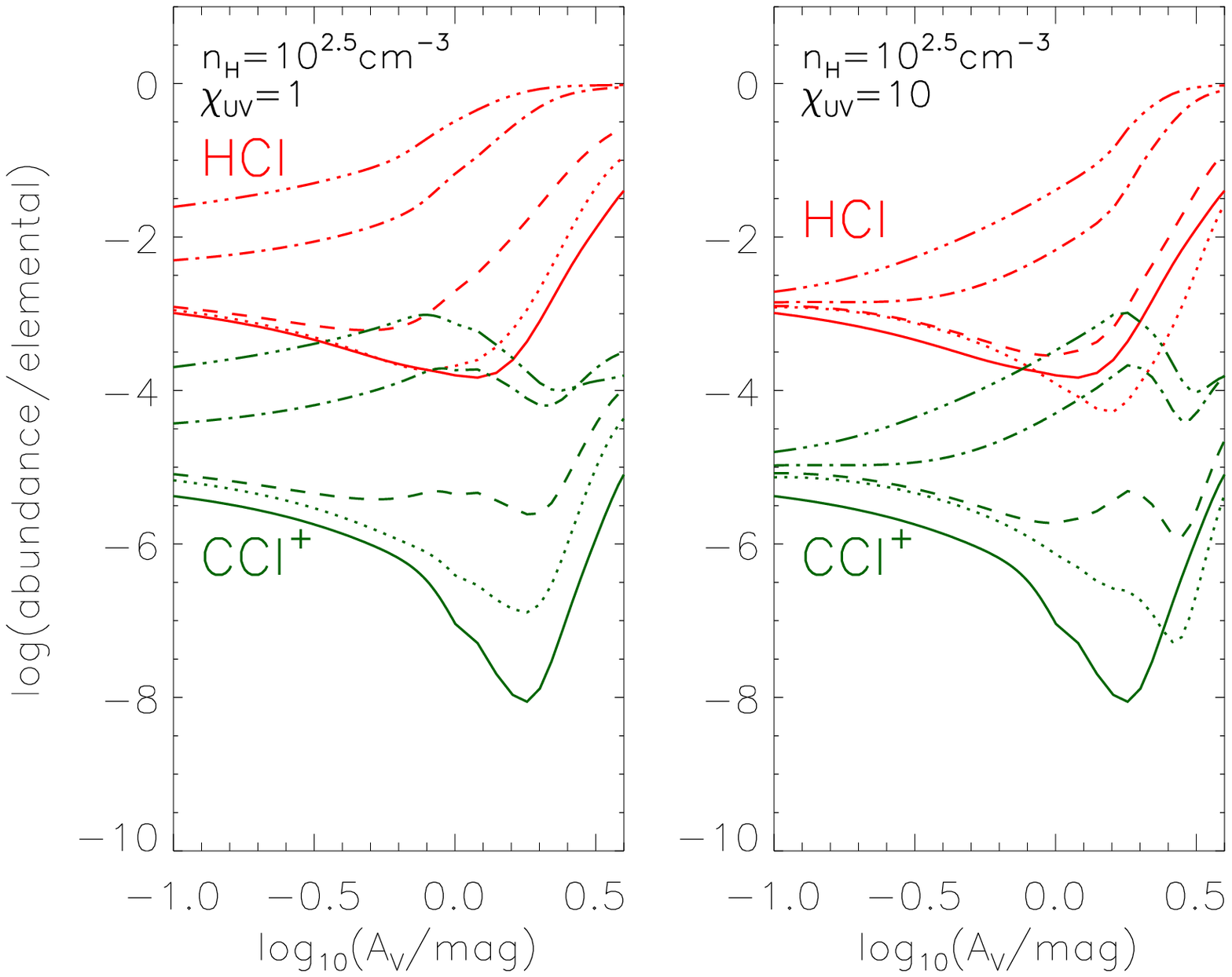}

\noindent{Fig.\ 8 --  Effect of enhanced temperatures on the abundances of HCl (red curves) and CCl$^+$ (green curves).  Results of one-sided models for slabs of density $n_{\rm H} = 10^{2.5} \, \rm \, cm^{-3}$ exposed to radiation with $\chi_{UV}$ = 1 (left panel) and 10 (right panel).  In addition to the standard model in which the temperature is determined by considerations of thermal balance (solid curve in each set), we show the abundance profiles for models in which the temperature is arbitrarily set to 150 K (dotted), 200 K (dashed) , 300 K (dot-dash), and 400 K (triple-dot-dash)}
\end{figure}
\begin{figure}
\includegraphics[scale=0.8,angle=0]{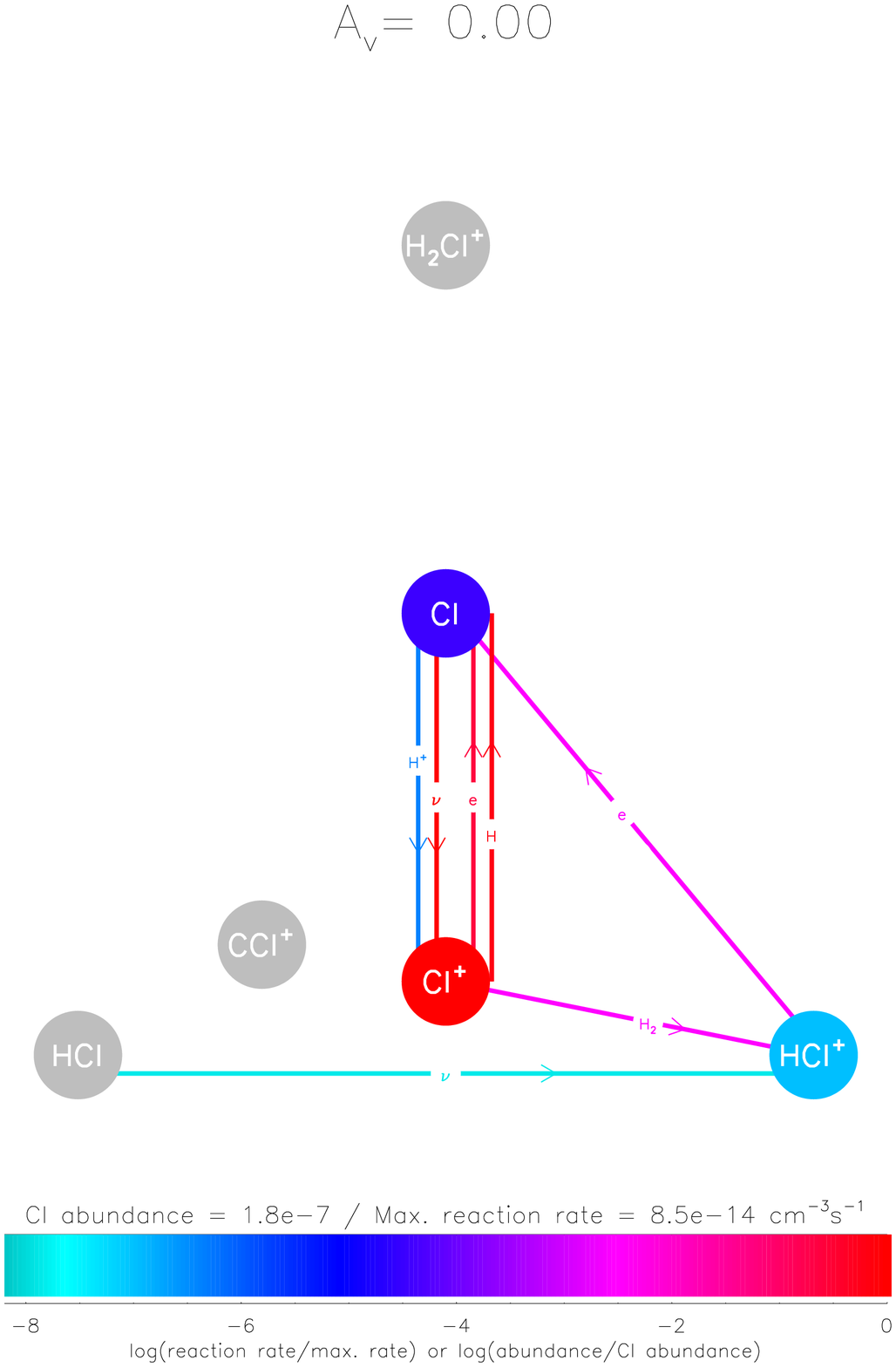}
\noindent{Fig. 9 --  Chemical network diagram, for a one-sided slab model with $\rm n_{\it H}=10^4\, cm^{-3}$ and $\chi_{UV}=10^4$.  The figure is labeled by the depth, $A_V$, in visual magnitude.}
\end{figure}
\begin{figure}
\includegraphics[scale=0.8,angle=0]{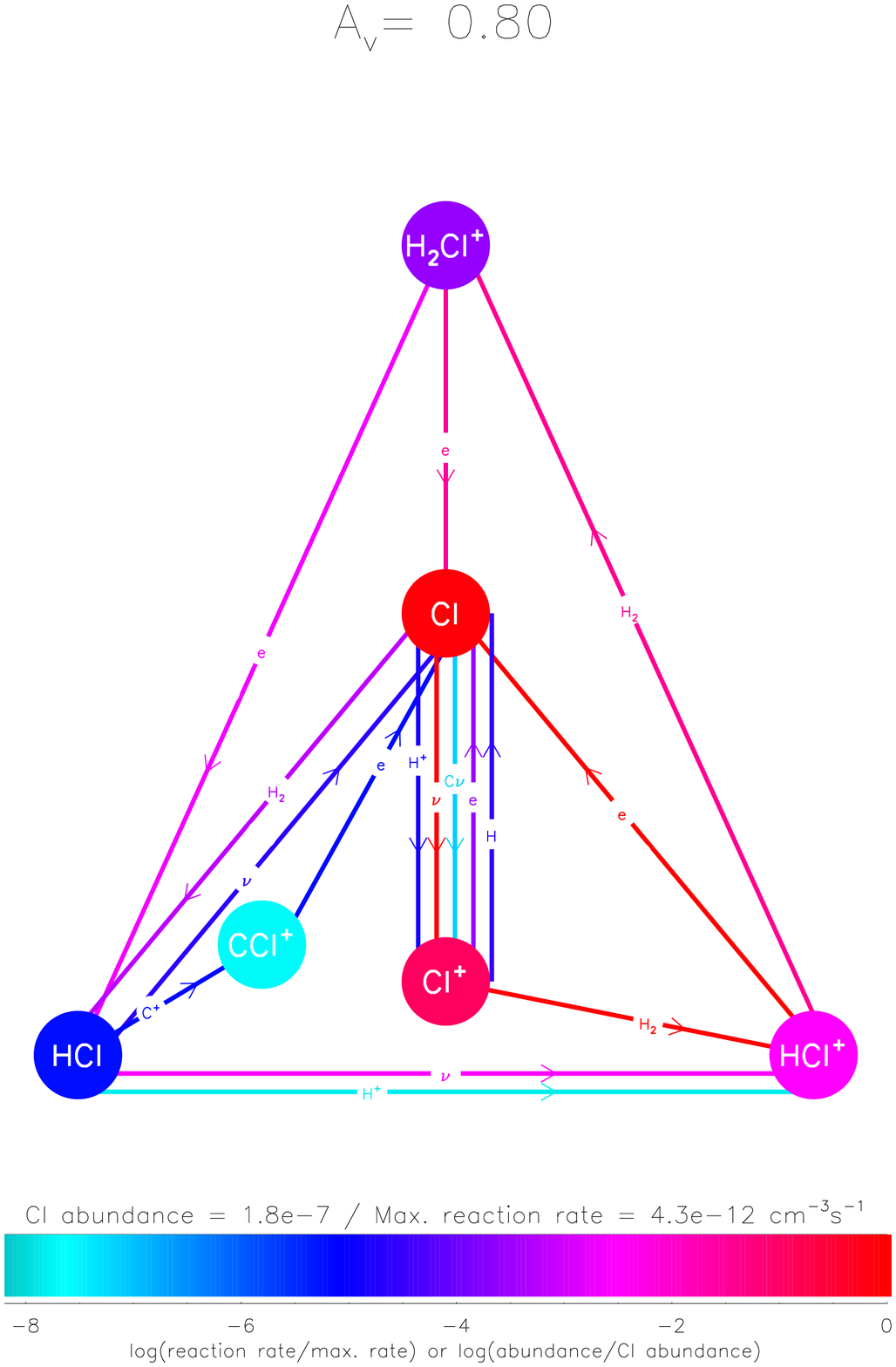}
\noindent{Fig. 10 --  Chemical network diagram, for a one-sided slab model with $\rm n_{\it H}=10^4\, cm^{-3}$ and $\chi_{UV}=10^4$.  The figure is labeled by the depth, $A_V$, in visual magnitude.}
\end{figure}
\begin{figure}
\includegraphics[scale=0.8,angle=0]{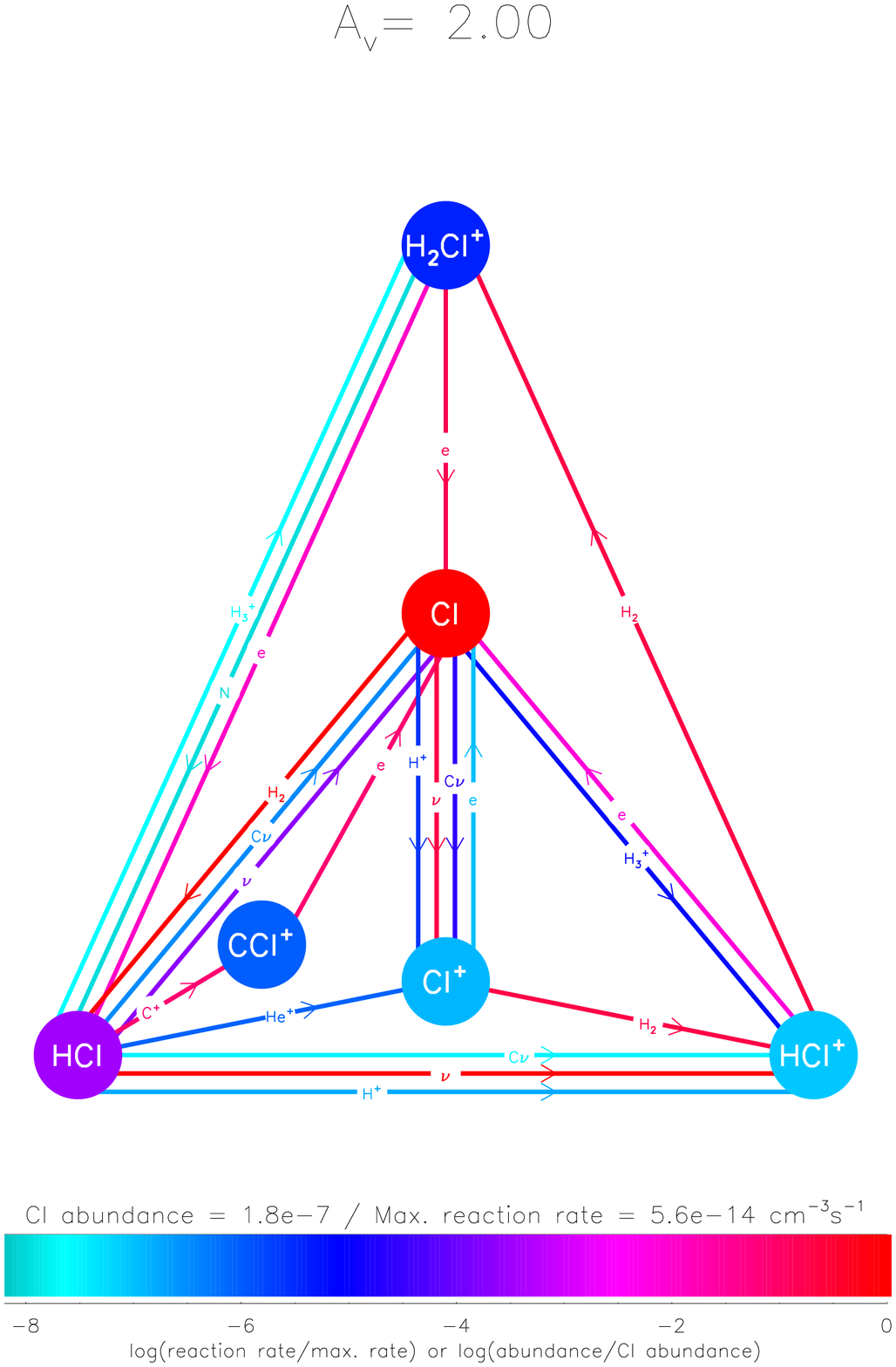}
\noindent{Fig. 11 --  Chemical network diagram, for a one-sided slab model with $\rm n_{\it H}=10^4\, cm^{-3}$ and $\chi_{UV}=10^4$.  The figure is labeled by the depth, $A_V$, in visual magnitude.}
\end{figure}
\begin{figure}
\includegraphics[scale=0.8,angle=0]{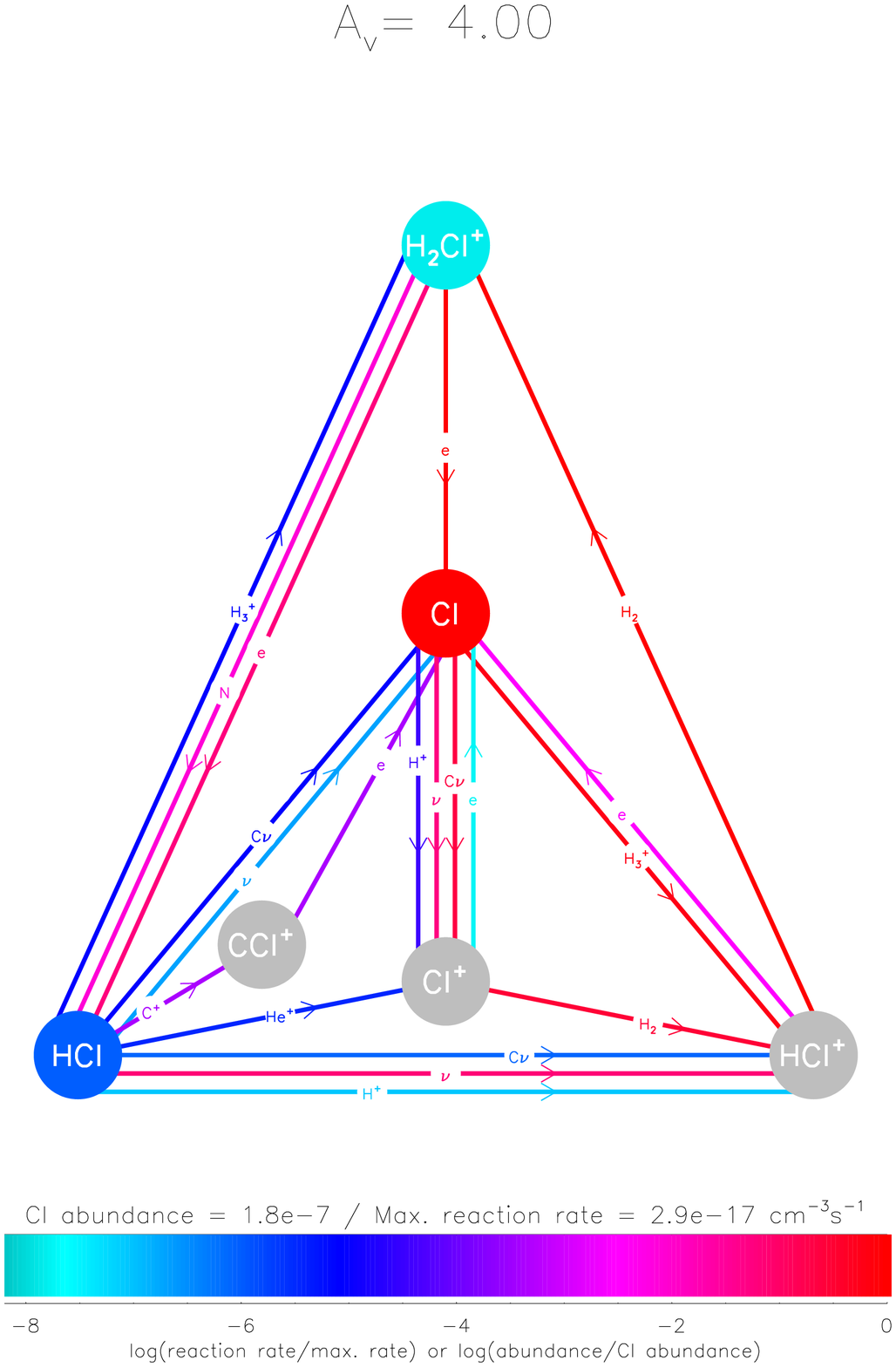}
\noindent{Fig. 12 --  Chemical network diagram, for a one-sided slab model with $\rm n_{\it H}=10^4\, cm^{-3}$ and $\chi_{UV}=10^4$.  The figure is labeled by the depth, $A_V$, in visual magnitude.}
\end{figure}
\begin{figure}
\includegraphics[scale=0.8,angle=0]{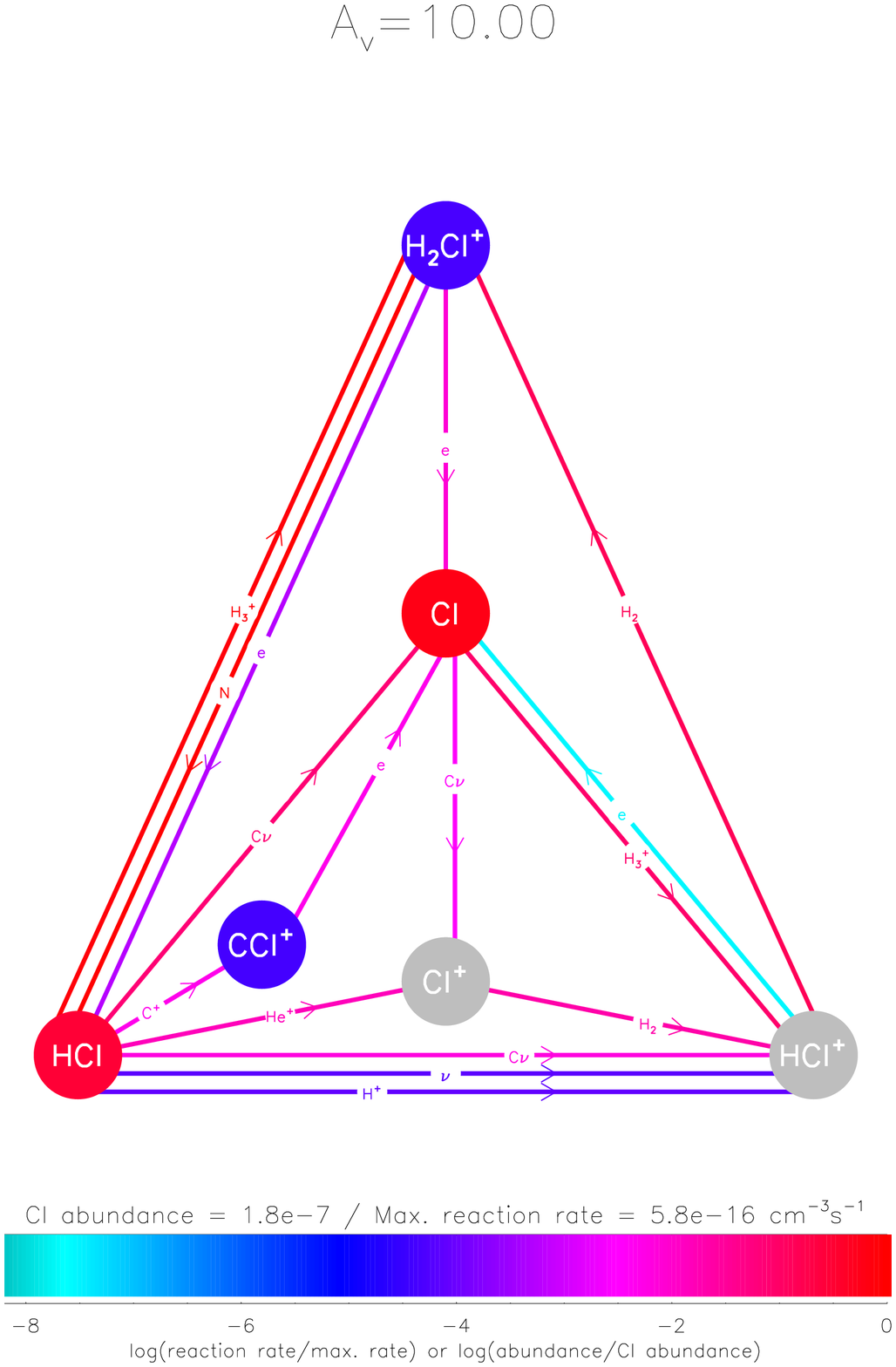}
\noindent{Fig. 13 --  Chemical network diagram, for a one-sided slab model with $\rm n_{\it H}=10^4\, cm^{-3}$ and $\chi_{UV}=10^4$.  The figure is labeled by the depth, $A_V$, in visual magnitude.}
\end{figure}
\begin{figure}
\includegraphics[scale=0.9,angle=0]{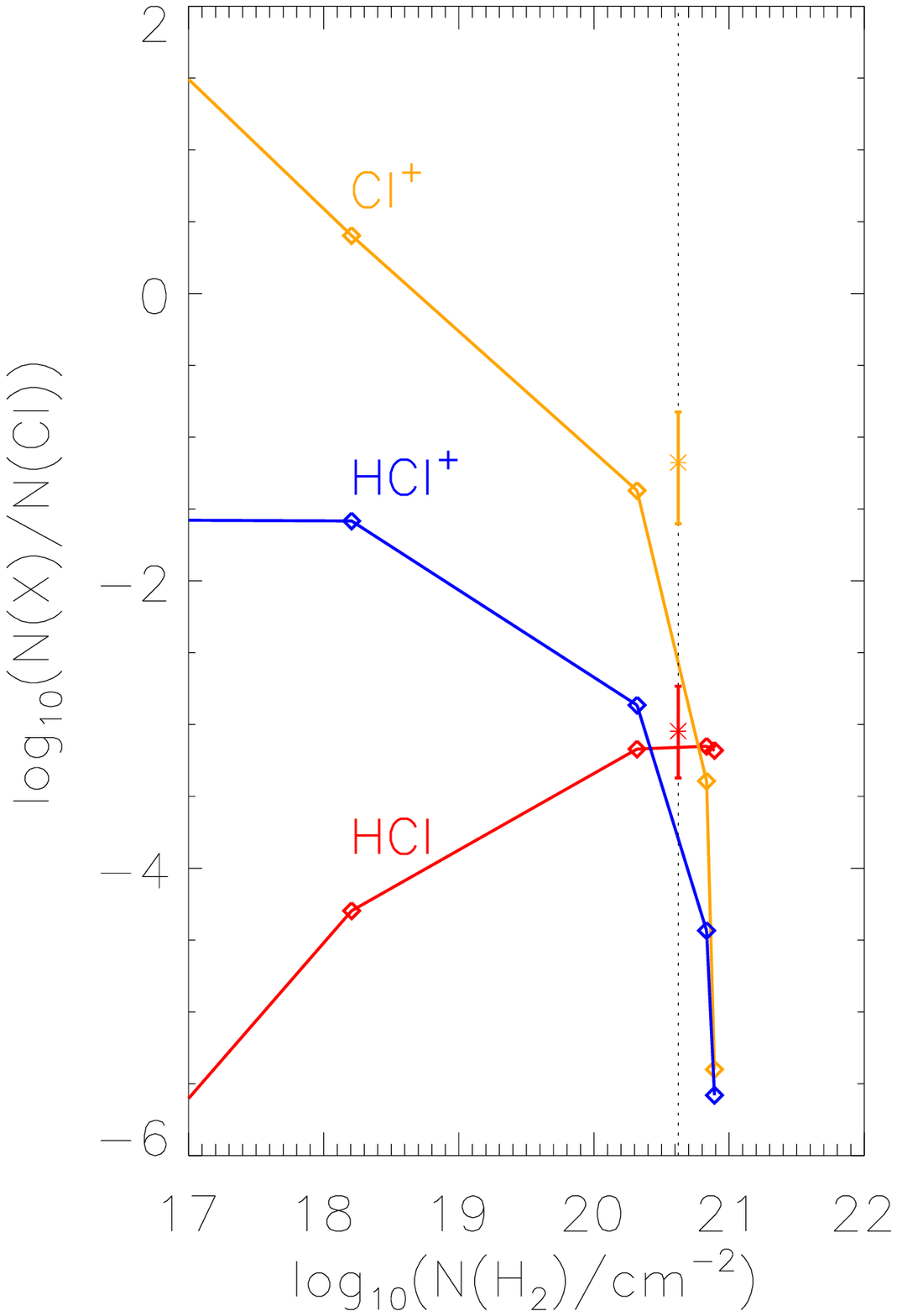}

\noindent{Fig.\ 14 -- Predictions from a series of models with $\chi_{UV}$ ranging from $10^{-1}$ to 10$^2$.  All models apply to a slab with parameters appropriate to $\zeta$~Oph, viz.\ density $n_H = 10^{2.5} \rm \, cm^{-3}$ and total visual extinction of 0.8 mag.  The horizontal axis shows the H$_2$ column density, for which the measured value is $4.2 \times 10^{20}$ along this sight-line, while the vertical axis shows $N({\rm HCl})/N({\rm Cl})$ (red curve), $N({\rm Cl^+})/N({\rm Cl})$ (orange) and $N({\rm HCl^+})/N({\rm Cl})$ (blue).  Squares located along the curves show the results for $\chi_{UV} = 10^{-1}$, 1, 10 and 10$^2$ from right to left. }
\end{figure}

\begin{thebibliography}{}


\bibitem[Amin(1996)]{1996EM&P...73..133A} Amin, M.~Y.\ 1996, Earth Moon and Planets, 73, 133 

\bibitem[Andric et al.(2002)]{2002PhLA..298...41A} Andric, L., Baccarelli, 
I., Grozdanov, T.~P., \& McCarroll, R.\ 2002, Physics Letters A, 298, 41 

\bibitem[Araki et al.(2001)]{2001JMoSp.210..132A} Araki, M., Furuya, T., 
\& Saito, S.\ 2001, Journal of Molecular Spectroscopy, 210, 132 

\bibitem[Armen]{Armen}Armentrout, P.~B., \& Sunderlin, L. S.\ 1992, Transition Metal Hydrides, ed.\ A.~Dedieu (VCH, New York, 1992)

\bibitem[Badnell et 
al.(2003)]{2003A&A...406.1151B} Badnell, N.~R., et al.\ 2003, \aap, 406, 1151 

\bibitem[Badnell(2006)]{2006ApJS..167..334B} Badnell, N.~R.\ 2006, \apjs, 
167, 334 

\bibitem[Blake et al.(1986)]{1986ApJ...300..415B} Blake, G.~A., Anicich, 
V.~G., \& Huntress, W.~T., Jr.\ 1986, \apj, 300, 415 (BAH86)

\bibitem[Blake et al.(1985)]{1985ApJ...295..501B} Blake, G.~A., Keene, J., 
\& Phillips, T.~G.\ 1985, \apj, 295, 501 

\bibitem[BC]{BC}Binning, R.~C., Jr., \& Curtiss, L. A.\ 1990, \jcp, 92, 1860

\bibitem[Brion et al.]{Brion} Brion, C.~E., Dyck, M., \& Cooper, G.\ 2005, Journal of Elect. Spectr. 144-147, 127

\bibitem[Brown et al.(1980)]{1980PhRvA..21.1237B} Brown, E.~R., Carter, 
S.~L., \& Kelly, H.~P.\ 1980, \pra, 21, 1237 

\bibitem[Chan et al.(1991)]{1991PhRvA..44..186C} Chan, W.~F., Cooper, G., 
\& Brion, C.~E.\ 1991, \pra, 44, 186 

\bibitem[Chen et al.(1991)]{1991JChPh..95.1228C} Chen, Y.-M., Clemmer, Huber, Κ. P., Herzberg, G.1979, Constants of Diatomic Molecules, Van Nostrand, New York.
D.~E., \& Armentrout, P.~B.\ 1991, \jcp, 95, 1228 

\bibitem[Chen et al.(1993)]{1993JChPh..98.4929C} Chen, Y.-M., Clemmer, 
D.~E., \& Armentrout, P.~B.\ 1993, \jcp, 98, 4929 

\bibitem[Cheng et al.(2002)]{2002JChPh.117.4293C} Cheng, B.-M., Chung, 
C.-Y., Bahou, M., Lee, Y.-P., \& Lee, L.~C.\ 2002, \jcp, 117, 4293 

\bibitem[Dalgarno et al.(1974)]{1974ApJ...192L..37D} Dalgarno, A., de Jong, 
T., Oppenheimer, M., \& Black, J.~H.\ 1974, \apjl, 192, L37 

\bibitem[Draine(1978)]{draine78} Draine, B.~T.\ 1978, ApJS, 36, 595 

\bibitem[Federman et al.(1995)]{1995ApJ...445..325F} Federman, S.~R., 
Cardell, J.~A., van Dishoeck, E.~F., Lambert, D.~L., 
\& Black, J.~H.\ 1995, \apj, 445, 325 

\bibitem[G]{G}Geppert, W. D., \& Hamberg, M.\ 2009, private communication.

\bibitem[HH]{HH}Huber, K. P., \& Herzberg, G. 1979, Constants of Diatomic Molecules (Van Nostrand, New York)

\bibitem[Indriolo et al.(2007)]{2007ApJ...671.1736I} Indriolo, N., Geballe, 
T.~R., Oka, T., \& McCall, B.~J.\ 2007, \apj, 671, 1736 

\bibitem[Jura(1974)]{1974ApJ...190L..33J} Jura, M.\ 1974, \apjl, 190, L33 

\bibitem[Kaufman et al.(1999)Kaufman, Wolfire, Hollenbach, \& 
   Luhman]{kauf99} Kaufman, M.~J., Wolfire, M.~G., 
   Hollenbach, D.~J., \& Luhman, M.~L.\ 1999, \apj, 527, 795 (K99)

\bibitem[Kaufman et al.(2006)]{2006ApJ...644..283K} Kaufman, M.~J.,
Wolfire, M.~G., \& Hollenbach, D.~J.\ 2006, \apj, 644, 283

\bibitem[KLM]{KLM} Kumaran, S.~S., Lim, K.~P., \& Michael, J.~V.\ 1994, \jcp, 101, 9487

\bibitem[Le Bourlot et al.(1993)Le Bourlot, Pineau Des Forets, Roueff, \& 
Flower]{lebour93} Le Bourlot, J., Pineau Des Forets, G., 
Roueff, E., \& Flower, D.~R.\ 1993, \aap, 267, 233 

\bibitem[Le Petit et al.(2006)]{2006ApJS..164..506L} Le Petit, F.,
Nehm{\'e}, C., Le Bourlot, J., \& Roueff, E.\ 2006, \apjs, 164, 506

\bibitem[Li 
\& Draine(2001)]{2001ApJ...554..778L} Li, A., \& Draine, B.~T.\ 2001, \apj, 554, 778 

\bibitem[Lubic et al.(1989)]{1989JMoSp.134....1L} Lubic, K.~G., Ray, D., 
Hovde, D.~C., Veseth, L., 
\& Saykally, R.~J.\ 1989, Journal of Molecular Spectroscopy, 134, 1 

\bibitem[Manthe et al.(2004)]{2004PCCP....6.5026M} Manthe, U., Capecchi, 
G., 
\& Werner, H.-J.\ 2004, Physical Chemistry Chemical Physics (Incorporating Faraday Transactions), 6, 5026 

\bibitem[Mazzotta et 
al.(1998)]{1998A&AS..133..403M} Mazzotta, P., Mazzitelli, G., Colafrancesco, S., \& Vittorio, N.\ 1998, \aaps, 133, 403 
200
\bibitem[Nee et al.(1986)]{1986JChPh..85..719N} Nee, J.~B., Suto, M., 
\& Lee, L.~C.\ 1986, \jcp, 85, 719 

\bibitem[Neufeld et al.(2005)]{2005ApJ...628..260N} Neufeld, D.~A., 
Wolfire, M.~G., \& Schilke, P.\ 2005, \apj, 628, 260 (NWS05)

\bibitem[Neufeld et 
al.(2006)]{2006A&A...454L..37N} Neufeld, D.~A., et al.\ 2006, \aap, 454, L37 

\bibitem[Novotny et al.(2005)]{2005JPhB...38.1471N} Novotny, O., et al.\ 
2005, Journal of Physics B Atomic Molecular Physics, 38, 1471 

\bibitem[Ph]{Ph} Phillips, T. G.\ et al.\ 2009, in preparation.

\bibitem[Pradhan et al.(1991)]{1991JChPh..95.9009P} Pradhan, A.~D., Kirby, 
K.~P., \& Dalgarno, A.\ 1991, \jcp, 95, 9009 

\bibitem[Pradhan 
\& Dalgarno(1994)]{1994PhRvA..49..960P} Pradhan, A., \& Dalgarno, A.\ 1994, \pra, 49, 960 

\bibitem[Ruscic 
\& Berkowitz(1983)]{1983PhRvL..50..675R} Ruscic, B., \& Berkowitz, J.\ 1983, Physical Review Letters, 50, 675 

\bibitem[Salez et al.(1996)]{1996ApJ...467..708S} Salez, M., Frerking, 
M.~A., \& Langer, W.~D.\ 1996, \apj, 467, 708 

\bibitem[Schilke et al.(1995)]{1995ApJ...441..334S} Schilke, P., Phillips, 
T.~G., \& Wang, N.\ 1995, \apj, 441, 334 

\bibitem[Sternberg et al. 1987]{ste87} Sternberg, A., Dalgarno, A., \& Lepp, S.\ 1987, ApJ, 320, 676

\bibitem[Stevens et al. (1989)]{ste89} Stevens, P.S., Brune, W.H., \& Anderson, J.G. 1989, J.\ Phys.\ Chem., 93, 4068 

\bibitem[Sonnentrucker et al.(2002)]{2002ApJ...576..241S} Sonnentrucker, 
P., Friedman, S.~D., Welty, D.~E., York, D.~G., 
\& Snow, T.~P.\ 2002, \apj, 576, 241 

\bibitem[Sonnentrucker et al.(2006)]{2006ApJ...650L.115S} Sonnentrucker, 
P., Friedman, S.~D., \& York, D.~G.\ 2006, \apjl, 650, L115 

\bibitem[Tielens \& Hollenbach(1985)]{th85} Tielens, 
A.~G.~G.~M.~\& Hollenbach, D.\ 1985, \apj, 291, 722 

\bibitem[Troe (1985)]{tro85} Troe, J.\ 1985, Chem.\ Phys.\ Lett., 122, 425

\bibitem[Troe (1987)]{tro87} Troe, J.\ 1987, J.\ Chem.\ Phys., 87, 2773

\bibitem[Troe (1996)]{tro96} Troe, J.\ 1996, J.\ Chem.\ Phys., 105, 6249

\bibitem[Weingartner \& Draine(2001)]{wd01} Weingartner, 
J.~C.~\& Draine, B.~T.\ 2001, \apjs, 134, 263 

\bibitem[van Dishoeck et al.(1982)]{1982JChPh..77.3693V} van Dishoeck, 
E.~F., van Hemert, M.~C., \& Dalgarno, A.\ 1982, \jcp, 77, 3693 (vDvHD)

\bibitem[van Dishoeck 
\& Black(1986)]{1986ApJS...62..109V} van Dishoeck, E.~F., \& Black, J.~H.\ 1986, \apjs, 62, 109 (vDB86)

\bibitem[van Dishoeck(1988)]{1988rcia.conf...49V} van Dishoeck, E.~F.\ 
1988, Rate Coefficients in Astrochemistry.~Proceedings of a Conference held 
in UMIST, Manchester, United Kingdom, September 21-24, 1987.~Editors, 
T.J.~Millar, D.A.~Williams; Publisher, Kluwer Academic Publishers, 
Dordrecht, Boston, 1988.~ISBN \# 90-277-2752-X.~LC \# QB450 .R38 
1988.~P.~49, 1988, 49 

\bibitem[Walmsley et al.(2002)]{2002ApJ...566L.109W} Walmsley, C.~M., 
Bachiller, R., Pineau des For{\^e}ts, G., 
\& Schilke, P.\ 2002, \apjl, 566, L109 

\bibitem[Wolfire, Tielens, \& Hollenbach(1990)]{wolfire90} 
Wolfire, M.~G., Tielens, A.~G.~G.~M., \& Hollenbach, D.\ 1990, \apj, 358, 116

\bibitem[Woodall et 
al.(2007)]{2007A&A...466.1197W} Woodall, J., Ag{\'u}ndez, M., Markwick-Kemper, A.~J., \& Millar, T.~J.\ 2007, \aap, 466, 1197 

\bibitem[Ziurys(1996)]{1996IAUS..170..370Z} Ziurys, L.~M.\ 1996, CO: 
Twenty-Five Years of Millimeter-Wave Spectroscopy, 170, 370 	 			

\bibitem[Zmuidzinas et al.(1995)]{1995ApJ...447L.125Z} Zmuidzinas, J., 
Blake, G.~A., Carlstrom, J., Keene, J., 
\& Miller, D.\ 1995, \apjl, 447, L125 

\end{thebibliography}
\end{document}